\documentclass[12pt]{iopart}

\usepackage{color}
\begin{document}

\title{Kinetic equations for systems with long-range interactions: a unified description}

\author{Pierre-Henri Chavanis}

\address{Laboratoire de Physique Th\'eorique,
Universit\'e Paul Sabatier,
118 route de Narbonne 31062 Toulouse, France}
\ead{chavanis@irsamc.ups-tlse.fr}
\begin{abstract}
We complete the existing literature on the kinetic theory of systems
with long-range interactions. Starting from the BBGKY hierarchy, or
using projection operator technics or a quasilinear theory, a general
kinetic equation can be derived when collective effects are
neglected. This equation (which is not well-known) applies to possibly
spatially {\it inhomogeneous} systems, which is specific to systems with
long-range interactions.  Interestingly, the structure of this kinetic
equation bears a clear physical meaning in terms of generalized Kubo
relations. Furthermore, this equation takes a very similar form for
stellar systems and two-dimensional point vortices providing therefore
a unified description of the kinetic theory of these systems. If we
assume that the system is spatially homogeneous (or axisymmetric for
point vortices), this equation can be simplified and reduces to the
Landau equation (or its counterpart for point vortices). Our formalism
thus offers a simple derivation of Landau-type equations. We also use
this general formalism to derive a kinetic equation, written in
angle-action variables, describing spatially inhomogeneous systems
with long-range interactions. This new derivation solves the
shortcomings of our previous derivation [P.H. Chavanis, Physica A {\bf
377}, 469 (2007)].  Finally, we consider a test particle approach and
derive general expressions for the diffusion and friction (or drift)
coefficients of a test particle evolving in a bath of field
particles. We make contact with the expressions previously obtained in
the literature. As an application of the kinetic theory, we argue
that the relaxation time is shorter for inhomogeneous (or high-dimensional)
systems than for homogeneous (or low-dimensional) systems because there
are potentially more resonances. We compare this prediction with existing numerical
results for the HMF model and 2D point vortices. For the HMF model,
we argue that the relaxation time scales as $N$ for
inhomogeneous distributions and  as $e^N$ for permanently
homogeneous distributions. Phase space structures
can reduce the relaxation time by creating some inhomogeneities and resonances.
Similar results are expected for 2D point vortices.

\end{abstract}

\maketitle

\section{Introduction}

Kinetic theories of systems with long-range interactions are important
to understand the dynamical evolution of the system and to study
transport properties. They can be used to determine the timescale of
``collisional'' relaxation towards the Boltzmann distribution, in
particular its dependence on the number of particles $N$. Furthermore,
they can tell whether or not the system will truly relax towards
statistical equilibrium. Indeed, it is not obvious that complex
systems (like systems with long-range interactions) do eventually
relax towards Boltzmann equilibrium because the ergodicity
assumption which sustains the statistical theory can break down.
Different kinetic theories have been developed over the years.

{\it Dilute gases:} The first kinetic theory of many-particles system
was elaborated by Boltzmann \cite{boltzmann} for a dilute gas. In that
case, the particles do not interact except during strong
collisions. Boltzmann derived his famous equation from which he proved
that the entropy increases ($H$-theorem) and that the system
ultimately relaxes towards the Maxwell distribution of
statistical equilibrium.

{\it Coulombian plasmas:} Landau \cite{landau} derived a kinetic
equation for a Coulombian plasma by starting from
the Boltzmann equation and considering a weak deflection
approximation. Indeed, for a Coulombian potential of interaction slowly
decreasing with the distance as $r^{-1}$, weak collisions are the most
frequent ones. Each encounter induces a {\it small} change in the
velocity of a particle but the cumulated effect of these encounters
leads to a macroscopic process of diffusion in velocity space. The
treatment of Landau, which assumes that the particles follow linear trajectories
with constant velocity in a first approximation, yields a logarithmic
divergence of the diffusion coefficient for both small and large
impact parameters but the equation can still be used successfully if
appropriate cut-offs are introduced. A natural lower cut-off, which is
called the Landau length, corresponds to the impact parameter leading
to a deflection at $90^{o}$. On the other hand, in a neutral plasma,
the potential is screened on a distance corresponding to the Debye
length. Phenomenologically, the Debye length provides an upper
cut-off. Later on, Lenard \cite{lenard} and Balescu
\cite{balescu} developed a more precise  kinetic
theory that could take into account collective effects. This gives
rise to the inclusion of the dielectric function $|\epsilon({\bf
k},{\bf k}\cdot{\bf v})|^{2}$ in the denominator of the potential of
interaction appearing in the kinetic equation. Physically, this means
that the particles are ``dressed'' by a polarization cloud. The
original Landau equation, which ignores collective effects, is
recovered from the Lenard-Balescu equation when $|\epsilon({\bf
k},{\bf k}\cdot{\bf v})|^{2}=1$. However, with this additional term,
it is found that the logarithmic divergence at large scales is now
removed and that the Debye length is indeed the natural upper
lengthscale to consider. At about the same period, Hubbard
\cite{hubbard} developed a test particle approach and derived a
Fokker-Planck equation describing the relaxation of a test particle in
a bath of field particles. He calculated the diffusion and friction
coefficients by evaluating the first and second moments of the
velocity deflection and took into account collective effects.  It is
interesting to note, for historical reasons, that Hubbard
\cite{hubbard} was apparently not aware of the works of Lenard
\cite{lenard} and Balescu \cite{balescu} at the same period and
that he developed his approach independently. However, the three
approaches are closely related. Indeed, if we substitute the diffusion
and friction coefficients found by Hubbard in the general form of the
Fokker-Planck equation and perform minor transformations (a
substitution that Hubbard has not explicitly made), one obtains the
Lenard-Balescu equation!

{\it Stellar systems:} In stellar dynamics, Chandrasekhar
\cite{chandra,chandra1,chandra2} developed a kinetic theory of stars in order to determine the
timescale of collisional relaxation and the rate of escape of stars
from globular clusters. To simplify the kinetic theory, he considered
an infinite and homogeneous system. He started from the general
Fokker-Planck equation and determined the diffusion coefficient and
the friction force (second and first moments of the velocity
increments) by considering the mean effect of a succession of two-body
encounters\footnote{Later, Rosenbluth {\it et al.} \cite{rosen}
proposed a simplified derivation of the coefficients of diffusion and
friction for stellar systems and for plasmas (without collective
effects).}. Since his approach can take into account large
deflections, there is no divergence at small impact parameters and the
gravitational analogue of the Landau length appears naturally in the
treatment of Chandrasekhar.  However, his approach leads to a
logarithmic divergence at large scales that is more difficult to
remove in stellar dynamics than in plasma physics because of the
absence of Debye shielding for the gravitational force. In a series of
papers, Chandrasekhar \& von Neumann \cite{cn} developed a completely
stochastic formalism of gravitational fluctuations and showed that the
fluctuations of the gravitational force are given by the Holtzmark
distribution (a particular L\'evy law) in which the nearest neighbor
plays a prevalent role. From these results, they argued that the
logarithmic divergence has to be cut-off at the interparticle
distance. However, since the interparticle distance is smaller than
the Debye length, the same arguments should also apply in plasma
physics, which is not the case. Therefore, the conclusions of
Chandrasekhar \& von Neumann are usually taken with circumspection. In
particular, Cohen {\it et al.} \cite{cohen} argue that the logarithmic
divergence should be cut-off at the physical size $R$ of the cluster
(or at the Jeans length) since the Jeans length is the presumable
analogue of the Debye length in the present context. These kinetic
theories lead to a relaxation time of the form $t_{R}\sim (N/\ln
N) t_D$ where $t_D$ is the dynamical time.  Chandrasekhar \cite{nice}
also developed a Brownian theory of stellar dynamics and showed that,
on a qualitative point of view, the results of kinetic theory can be
understood very simply in that framework. In particular, he showed
that the {\it dynamical friction} is necessary to reproduce the
Maxwell-Boltzmann distribution at equilibrium and that the
coefficients of diffusion and friction are related to each other by an
Einstein relation. This relation is confirmed by his more precise
kinetic theory.  It is important to emphasize, however, that
Chandrasekhar did not derive the kinetic equation for the evolution of
the system {\it as a whole}. Indeed, he considered the Brownian motion
of a test star in a {\it fixed} distribution of field stars (bath) and
derived the corresponding Fokker-Planck equation\footnote{Indeed,
Chandrasekhar
\cite{chandra1,chandra2} models the evolution of
globular clusters by the Kramers equation which has a fixed
temperature (canonical description) while a more relevant kinetic
equation would be the Landau equation which conserves the energy
(microcanonical description).}. This equation has been used to study
the evaporation of stars from globular clusters
\cite{chandra2,spitzer,michie,king,lcstar}. King \cite{kingL} noted
that, if we were to describe the dynamical evolution of the cluster as
a whole, the distribution of the field particles must evolve in time
in a self-consistent manner so that the kinetic equation must be an
integrodifferential equation. The kinetic equation obtained by King is
equivalent to the Landau equation\footnote{This is not obvious on the
form given by King, that is deduced from the work of Rosenbluth {\it
et al.} \cite{rosen}. However, the connection with the Landau equation
can be made easily by using simple integrations by parts; see Ref. \cite{epjb}
for a recent discussion of that issue.}. It is interesting
to note, for historical reasons, that none of the previous authors
seemed to be aware of the work of Landau \cite{landau} at that
time\footnote{To our knowledge, the first explicit reference to the
Landau equation appeared much later in the paper of Kandrup
\cite{kandrup1}. There is also a strange comment related to the work
of Landau in the paper of Cohen {\it et al.}
\cite{cohen}.}. There is,
however, an important difference between stellar dynamics and plasma
physics. Neutral plasmas are usually spatially homogeneous due to
Debye shielding.  By contrast, stellar systems are spatially
inhomogeneous. The above-mentioned kinetic theories developed for an
infinite homogeneous system can be applied to an inhomogeneous system
only if we make a {\it local approximation}. In that case, the
collision term is calculated as if the system were spatially
homogeneous or as if the collisions could be treated as local. Then,
the effect of spatial inhomogeneity is only retained in the advective
(Vlasov) term which describes the evolution of the system due to
mean-field effects\footnote{The nowadays called ``Vlasov equation''
was introduced by Jeans \cite{jeans} in stellar dynamics and by Vlasov
\cite{vlasov} in plasma physics. See H\'enon \cite{henonvlasov} for
interesting comments about the name that one should give to that
equation.}. This leads to the Vlasov-Landau-Poisson system which is
the standard kinetic equation of stellar dynamics. To our knowledge,
this equation has been first written and studied by H\'enon
\cite{henon}. H\'enon also exploited the timescale separation between
the dynamical time $t_D$ and the relaxation time $t_{R}\gg t_D$ to
derive a simplified kinetic equation for $f(\epsilon,t)$, where
$\epsilon=v^2/2+\Phi({\bf r},t)$ is the individual energy, called the
orbit-averaged-Fokker-Planck equation. In this approach, the
distribution function $f({\bf r},{\bf v},t)$, averaged over a short
timescale, is a steady state of the Vlasov equation of the form
$f(\epsilon,t)$ which slowly evolves in time (on a long timescale) due
to the development of ``collisions'' (i.e. correlations due to finite
$N$ effects or graininess). Cohn \cite{cohn} numerically solved this
equation to describe the collisional evolution of star clusters and
investigate the gravothermal catastrophe that was predicted by Antonov
\cite{antonov} and Lynden-Bell \& Wood \cite{lbw} on the basis of
statistical mechanics. The local approximation, which is a crucial
step in the kinetic theory, is supported by the stochastic approach of
Chandrasekhar \& von Neumann
\cite{cn} showing the preponderance of the nearest
neighbor. However, this remains a simplifying assumption which is not
easily controllable. In particular, as we have already indicated, the
local approximation leads to a logarithmic divergence at large scales
that is difficult to remove. This divergence would not have occurred
if full account of spatial inhomogeneity had been given since the
start. On the other hand, the Fokker-Planck equation is based on a
Markov assumption and this assumption is not clearly justified for
stellar systems. Memory effects can be important for self-gravitating
systems because, as shown by Chandrasekhar \cite{ct}, the temporal
correlation function of the force decreases algebraically as $1/t$
instead of exponentially. This slow decay results in a logarithmic
divergence of the diffusion coefficient for large times when one
considers the Kubo formula. As shown by Lee \cite{lee}, the spatial
and temporal logarithmic divergences are equivalent, i.e. they are two
manifestations of the same phenomenon. The effect of spatial and
temporal delocalization was investigated by Gilbert \cite{gilbert},
Severne \& Haggerty \cite{severne}, and Kandrup \cite{kandrup1}. In
particular, Kandrup derived a generalized Landau equation by using
projection operator technics. Recently, Chavanis \cite{paper3,paper4}
obtained this equation in a simpler manner from the BBGKY hierarchy or
from a quasilinear theory as an expansion in $1/N$ in a proper
thermodynamic limit.  This generalized kinetic equation is interesting
because it can take into account effects of spatial inhomogeneity and
memory which are neglected in the previous approaches. It clearly
shows which approximations are needed in order to recover the Landau
equation. However, the generalized Landau equation remains extremely
complicated for practical applications.

{\it Two-dimensional point vortices and non-neutral plasmas:} It is
well-known that a non-neutral plasma under a strong magnetic field is
isomorphic to a system of point vortices in 2D hydrodynamics
\cite{mj}. These systems
are described by Hamiltonian equations in which the coordinates $x$
and $y$ of the particles are canonically conjugate
\cite{kirchhoff}. The particles (charges or vortices) interact through
a long-range potential that is played by the electric potential in a
plasma or by the stream function in 2D hydrodynamics. The statistical
mechanics of 2D point vortices was pioneered by Onsager \cite{onsager}
and further developed by Montgomery \& Joyce \cite{mj} and
Lundgren \& Pointin \cite{lp}  in a mean field approximation (recently,
Eyink \& Sreenivasan \cite{esr} have discovered that Onsager had first
performed this mean field theory in unpublished notes). Kinetic theories
of these systems were developed much later. Dubin \& O'Neil
\cite{dn}  derived a kinetic equation for a 2D guiding center plasma
by starting from the Klimontovich equation and using a quasilinear
theory. Independently, Chavanis \cite{kin} derived a kinetic equation
for 2D point vortices by using projection operator technics and, more
recently \cite{kinvortex}, by using a BBGKY-like hierarchy or a
quasilinear theory. The kinetic equation derived by Dubin \& O'Neil
takes into account collective effects and can be considered as the
counterpart of the Lenard-Balescu equation in plasma physics. The
kinetic equation derived by Chavanis ignores collective effects and
can be considered as the counterpart of the Landau equation in plasma
physics and stellar dynamics. However, the approach of Chavanis is
formulated for an arbitrary distribution of point vortices while the
approach of Dubin \& O'Neil is restricted to axisymmetric
distributions. These kinetic equations are valid at the order $O(1/N)$
in a proper thermodynamic limit, so that they describe the evolution
of the system on a timescale of order $Nt_D$. For
$N\rightarrow +\infty$, we get the 2D Euler equation which describes
collisionless systems. At the order $1/N$, the collision operator describes
the development of correlations between point vortices. However, when
we consider axisymmetric distributions, it is found that the collision
operator cancels out when the profile of angular velocity becomes
monotonic. In that case, the evolution stops even if the system has
not reached statistical equilibrium
\cite{kin,cl,kinvortex}. This implies that the relaxation time of the system as a
whole is larger than $Nt_D$, where $t_D$ is the dynamical time, so
that higher order correlations (three-body, four-body...) must be
advocated \cite{kin}. It is also possible that the point vortex gas
does not reach the Boltzmann distribution predicted
by statistical mechanics. Indeed, the dynamics may
be non-ergodic as discussed by Khanin \cite{khanin}.  On the other
hand, using an analogy with stellar dynamics and Brownian theory,
Chavanis
\cite{preR,kin,kinvortex} derived a Fokker-Planck equation describing
the relaxation of a test vortex in a bath of field vortices at
statistical equilibrium. The relaxation is due to a competition
between a diffusion term and a drift term. The {\it systematic drift} \cite{preR},
which is the counterpart of the dynamical friction \cite{nice} in stellar dynamics,
is necessary to obtain the Boltzmann distribution at statistical
equilibrium.  The diffusion coefficient and the drift term are related
to each other by an appropriate Einstein relation and {\it they are
inversely proportional to the local shear created by the field
vortices}. The properties of this Fokker-Planck equation have been
studied by Chavanis \& Lemou
\cite{cl}. On the other hand, Chavanis \& Sire \cite{cs} have studied the
statistics of the velocity fluctuations arising from a random
distribution of point vortices by using an approach similar to the one
developed by Chandrasekhar \& von Neumann
\cite{cn} for the gravitational force, but leading to different results due to the
lower dimension of space. The numerous analogies between the
statistical mechanics and the kinetic theories of stellar systems and 2D
vortices are discussed by Chavanis
\cite{houches}\footnote{As reported by Eyink \& Sreenivasan
\cite{esr}, Onsager first pointed out analogies between stellar
systems and 2D vortices.  In a Letter to Lin he wrote: ``At {\it
negative} temperatures, the appropriate statistical methods have
analogues not in the theory of electrolytes, but in the statistics of
stars...''. However, Onsager did not develop this matter
further.}. Other interesting kinetic theories of 2D point vortices
have been developed by Nazarenko \& Zakharov \cite{nz}, Marmanis
\cite{marmanis}, Sire \& Chavanis \cite{sc} and Newton \& Mezic
\cite{nm}.

{\it HMF model:} in the past, the kinetic theories of systems with
long-range interactions have been essentially developed for plasmas,
stellar systems and 2D vortices. Recently, there was a renewed
interest from a part of the community of statistical mechanics for
systems with long-range interactions. In that context, a simple model
of systems with long-range interactions, the so-called Hamiltonian
Mean Field (HMF) model, has been extensively studied \cite{cdr}. It
consists of $N$ particles moving on a ring and interacting via a
cosine potential. It can be viewed as a one dimensional periodic
system where the potential of interaction is truncated to one Fourier
mode. This is certainly the simplest system with long-range
interactions that we can imagine. However, it remains highly non
trivial and, interestingly, it exhibits many features common to other
systems with long-range interactions such as self-gravitating
systems. The kinetic theory of the spatially homogeneous phase of the
HMF model has been developed by Bouchet \cite{bouchet}, Bouchet \&
Dauxois \cite{bd} and Chavanis {\it et al.} \cite{cvb} (Inagaki
\cite{inagakikin} first developed a kinetic theory of the HMF model
but he arrived at incorrect conclusions). They considered an expansion
of the equations of the dynamics in powers of $1/N$ in a proper
thermodynamic limit. For $N\rightarrow +\infty$, this leads to the
Vlasov equation. At the order $O(1/N)$, this leads to the one
dimensional version of the Lenard-Balescu equation. However, the
Lenard-Balescu collision operator is known to vanish in one dimension
\cite{feix,kp}. This implies that the relaxation time is larger than
$N t_D$, where $t_D$ is the dynamical time. This result of kinetic
theory is in agreement with direct numerical simulations that lead to
$t_{R}\sim N^{1.7} t_D$
\cite{yamaguchi} or even $t_{R}\sim e^N t_D$ \cite{campa} depending on
the initial conditions. These authors also considered the relaxation
of a test particle in a bath of field particles. It is described by a
Fokker-Planck equation, involving a diffusion and a friction, which is
the one dimensional version of the Fokker-Planck equation derived by
Hubbard \cite{hubbard} in plasma physics. It can also be deduced from
the Lenard-Balescu equation by making a bath
approximation\footnote{The Lenard-Balescu equation and the
Fokker-Planck equation keep the same form in any dimension of space,
and for any potential of interaction $u(|{\bf r}-{\bf r}'|)$,
differing only in a multiplicative constant $(2\pi)^d$
\cite{epjb}. Therefore, they can be directly applied to the HMF model,
and more generally to any system with long-range
interactions. However, the explicit expressions of the diffusion
coefficient and friction force depend on the dimension of space. Some
useful expressions in $d=3,2,1$ dimensions are given in
\cite{epjb}.}. Bouchet \& Dauxois \cite{bd} used this Fokker-Planck
equation to show that the temporal correlation function of the
velocity decreases algebraically with time leading to (weak or strong)
anomalous diffusion (related results were previously obtained by
Marksteiner {\it et al.} \cite{marksteiner} in the very different
context of optical lattices).  The theoretical results of Bouchet \&
Dauxois \cite{bd} are in very good agreement with direct numerical
simulations in the situations considered \cite{yamabis}. Chavanis \&
Lemou \cite{front}, developing the theory of Potapenko {\it et al.} 
\cite{potapenko}, used the Fokker-Planck equation to study the
relaxation of the distribution function tail and show that it has a
front structure moving very slowly (logarithmically) with time.  Some
features of the kinetic theory of the spatially {\it inhomogeneous}
phase of the HMF model have been considered in Ref. \cite{curious}.

A first objective of the present paper is to develop a unified kinetic
theory of systems with long-range interactions. In particular, we show
that, when collective effects are neglected, a generalized kinetic
equation can be given for stellar systems [see Eq. (\ref{sw2})], for
the HMF model [see Eq. (\ref{sw2})] and for 2D point vortices [see
Eq. (\ref{vw2})]. This equation, which is not well-known, applies to
spatially inhomogeneous systems (or non axisymmetric distributions of
point vortices) and bears a clear physical meaning in terms of
generalized Kubo relations.  Furthermore, it avoids the use of
Fourier-Laplace transforms and remains in physical space. For
homogeneous distributions (or for axisymmetric distributions in the
case of point vortices), it reduces to the Landau equation (or to its
counterpart in 2D hydrodynamics). Therefore, our paper offers a new
derivation of Landau-type equations from a simple formalism and
develops the calculations in detail. A drawback of our approach is to
neglect collective effects that would lead to Lenard-Balescu-type
equations in the homogeneous case. A second objective of the paper is
to use this general formalism to derive a kinetic equation, written in
angle-action variables, describing spatially inhomogeneous systems
with long-range interactions. This new derivation solves the
shortcomings of our previous derivation
\cite{action}. We also derive general expressions for the diffusion
and friction (or drift) coefficients of a test particle evolving in a
bath of field particles and make contact with the expressions
previously obtained in the literature. Finally, we use this kinetic
theory to argue that the relaxation time is shorter for inhomogeneous
(or high-dimensional) systems than for homogeneous (or
low-dimensional) systems because there are potentially more
resonances, and we compare this prediction with existing numerical
results. For the HMF model, we argue that the
relaxation time scales as $N$ for inhomogeneous distributions and as
$e^N$ for permanently homogeneous distributions. Phase space
structures can reduce the relaxation time by creating some
inhomogeneities and resonances.  Similar results are expected for 2D
point vortices.

\section{Stellar systems and HMF model}
\label{sec_stellar}

\subsection{The evolution of the system as a whole: the kinetic equation}
\label{sec_sw}

Let us consider an isolated system of particles with long-range
interactions described by the Hamiltonian equations
\begin{eqnarray}
m\frac{d{\bf r}_i}{dt}=\frac{\partial H}{\partial {\bf v}_i},\qquad m\frac{d{\bf v}_i}{dt}=-\frac{\partial H}{\partial {\bf r}_i},\nonumber\\
H=\frac{1}{2}\sum_{i=1}^N m v_i^2+m^2\sum_{i<j}u({\bf r}_i-{\bf r}_j),
\label{sw1}
\end{eqnarray}
where $u({\bf r}-{\bf r}')$ is a binary potential of interaction
depending only on the absolute distance $|{\bf r}-{\bf r}'|$ between
the particles. We consider the proper thermodynamic limit
$N\rightarrow +\infty$ in such a way that the normalized energy
$\epsilon=E/(u_* N^2m^2)$ and the normalized temperature $\eta=\beta
Nm^2u_*$ are of order unity, where $u_*$ represents the typical value
of the potential of interaction \cite{assise}. By a suitable
normalization of the parameters, we can consider $N\rightarrow
+\infty$ with $V\sim E/N\sim\beta\sim m\sim 1$ and $u_*\sim 1/N$. This
is the usual Kac prescription which amounts to putting $\frac{1}{N}$
in front of the interaction energy (for rigorous mathematical results
see
\cite{ms,kiessling}). We can also consider $N\rightarrow +\infty$ with
$V\sim E\sim \beta/N\sim u_*\sim 1$ and $m\sim 1/N$. When collective
effects are ignored, the kinetic equation describing the evolution of
the distribution function $f({\bf r},{\bf v},t)$ at the order $O(1/N)$
is given by
\begin{eqnarray}
\frac{\partial f}{\partial t}+{\bf v}\cdot {\partial f\over\partial {\bf r}}+\frac{N-1}{N}\langle {\bf F}\rangle \cdot {\partial f\over \partial {\bf v}}=\frac{\partial}{\partial
{v}^{\mu}}\int_0^{t} d\tau \int d{\bf r}_{1}d{\bf v}_1
{F}^{\mu}(1\rightarrow
0)G(t,t-\tau)\nonumber\\
\times  \left \lbrack {{\cal F}}^{\nu}(1\rightarrow 0) {\partial\over\partial { v}^{\nu}}+{{\cal F}}^{\nu}(0\rightarrow 1) {\partial\over\partial {v}_{1}^{\nu}}\right \rbrack {f}({\bf r},{\bf v},t-\tau)\frac{f}{m}({\bf r}_1,{\bf
v}_1,t-\tau),\label{sw2}
\end{eqnarray}
where ${\bf F}(1\rightarrow 0)$ is the force (by unit of mass) created
by particle $1$ (with position ${\bf r}_{1}$ and velocity ${\bf v}_{1}$)
on particle $0$ (with position ${\bf r}$ and velocity ${\bf v}$)
and ${\cal F}^\mu(1\rightarrow 0)={\bf F}(1\rightarrow 0)-\langle {\bf
F}\rangle/N$ is the fluctuating force with respect to the average
force $\langle {\bf F}\rangle ({\bf r},t)=-\nabla\Phi({\bf r},t)$ experienced by particle
$0$. On the other hand, $G(t,t-\tau)$ is the Greenian constructed with
the mean flow in phase space.  The kinetic equation (\ref{sw2}) is not
well-known and this is why we try to publicize it in this paper. It
has been obtained by Kandrup \cite{kandrup1} from a projection
operator formalism and by Chavanis \cite{paper3,paper4} from the BBGKY
hierarchy and from a quasilinear theory. The BBGKY hierarchy is closed by
neglecting the three body correlation function which is of order
$O(1/N^2)$. The structure of this
equation bears a clear physical meaning. It involves a diffusion term
and a friction term. The coefficients of diffusion and
friction are given by generalized Kubo formulae, i.e. the time
integral of the temporal correlations of the fluctuating force. The
collision term $C_N[f]$ (r.h.s.) is valid at the order $1/N$
so that it describes the ``collisional'' evolution
of the system (ignoring collective effects) on a timescale of order
$Nt_D$. For $N\rightarrow +\infty$, we obtain the Vlasov equation in
which collisions (more properly correlations) are neglected (for rigorous
mathematical results see \cite{bh,hj,cr}). When
coupled to an attractive long-range potential (e.g., the gravitational
potential), the Vlasov equation can generate a process of violent collisionless
relaxation on a few dynamical times $t_D$. A statistical theory
of violent relaxation has been developed  by Lynden-Bell \cite{lb} for stellar systems,
but its domain of applicability is more general.
Equation (\ref{sw2}) is a
non-Markovian integro-differential equation. It can describe the
evolution of stellar systems, or the evolution of the HMF model,
taking into account delocalizations in space and time (i.e. spatial
inhomogeneity and memory effects). If we make a Markovian
approximation and extend the time integral to $+\infty$, we obtain
\begin{eqnarray}
\frac{\partial f}{\partial t}+{\bf v}\cdot {\partial f\over\partial {\bf r}}+\frac{N-1}{N}\langle {\bf F}\rangle \cdot {\partial f\over \partial {\bf v}}=\frac{\partial}{\partial
{v}^{\mu}}\int_0^{+\infty} d\tau \int d{\bf r}_{1}d{\bf v}_1
{F}^{\mu}(1\rightarrow
0)\nonumber\\
\times  G(t,t-\tau)\left \lbrack {{\cal F}}^{\nu}(1\rightarrow 0) {\partial\over\partial { v}^{\nu}}+{{\cal F}}^{\nu}(0\rightarrow 1) {\partial\over\partial {v}_{1}^{\nu}}\right \rbrack {f}({\bf r},{\bf v},t)\frac{f}{m}({\bf r}_1,{\bf
v}_1,t). \label{sw3}
\end{eqnarray}
The Markov approximation is not rigorously justified for
self-gravitating systems because the force auto-correlation function
decreases algebraically like $1/t$ \cite{ct}, instead of
exponentially. The Markov approximation is also expected to be
incorrect for the HMF model and other systems with long-range
interactions if we are close to the critical point. However, except
for these situations, the Markovian approximation should be justified
in the $N\rightarrow +\infty$ limit because the timescale $\sim Nt_D$
on which $f({\bf r},{\bf v},t)$ changes is long compared to the
timescale $\tau_{corr}$ for which the integrand in Eq. (\ref{sw2}) has
significant support.  The Markovian equation (\ref{sw3}) applies to
possibly spatially inhomogeneous distribution functions, which is a
specificity of systems with long-range interactions. Now, for
spatially homogeneous distribution functions $f({\bf v},t)$, the
kinetic equation reduces to
\begin{eqnarray}
\frac{\partial f}{\partial t}=\frac{\partial}{\partial
{v}^{\mu}}\int_0^{+\infty} d\tau \int d{\bf r}_{1}d{\bf v}_1
{F}^{\mu}(1\rightarrow
0)G(t,t-\tau)\nonumber\\
\times  \left \lbrack {{F}}^{\nu}(1\rightarrow 0) {\partial\over\partial { v}^{\nu}}+{{F}}^{\nu}(0\rightarrow 1) {\partial\over\partial {v}_{1}^{\nu}}\right \rbrack {f}({\bf v},t)\frac{f}{m}({\bf
v}_1,t). \label{sw4}
\end{eqnarray}
This is valid as long as the homogeneous distribution is Vlasov stable. In
fact, the distribution $f({\bf v},t)$ will change due to the
development of correlations and graininess effects (i.e. the r.h.s. of the
kinetic equation) and, at some point of the evolution, the
distribution may becomes Vlasov unstable. In that case, a dynamical
phase transition from a homogenous state to an inhomogeneous state
can take place. Such a transition has been illustrated numerically
for the HMF model by Campa {\it et al.} \cite{campat}.  Using the symmetry of
the force ${\bf F}(0\rightarrow 1)=-{\bf F}(1\rightarrow 0)$, and the
equations of the unperturbed trajectory
\begin{eqnarray}
{\bf v}(t-\tau)={\bf v}(t)={\bf v},\qquad {\bf r}(t-\tau)={\bf r}(t)-{\bf v}(t)\tau={\bf r}-{\bf v}\tau,
\label{sw5}
\end{eqnarray}
corresponding to  $\langle {\bf F}\rangle={\bf 0}$, the kinetic equation can be written
\begin{eqnarray}
\frac{\partial f}{\partial t}=\frac{\partial}{\partial
{v}^{\mu}}\int d{\bf v}_1 K^{\mu\nu}
 \left ({\partial\over\partial { v}^{\nu}}-{\partial\over\partial {v}_{1}^{\nu}}\right ) {f}({\bf v},t)\frac{f}{m}({\bf
v}_1,t), \label{sw6}
\end{eqnarray}
with
\begin{eqnarray}
K^{\mu\nu}=\int_0^{+\infty} d\tau \int d{\bf r}_{1}
{F}^{\mu}(1\rightarrow
0,t){{F}}^{\nu}(1\rightarrow 0,t-\tau). \label{sw7}
\end{eqnarray}
Now, the force (by unit of mass) created by particle $1$ on particle $0$ is given by
\begin{eqnarray}
{\bf F}(1\rightarrow 0)=-m\frac{\partial u}{\partial {\bf r}}({\bf r}-{\bf r}_{1}).\label{sw8}
\end{eqnarray}
Introducing the Fourier transform of the potential
\begin{eqnarray}
u({\bf x})=\int e^{i{\bf k}\cdot {\bf x}}\hat{u}({\bf k})\, d{\bf k},\qquad \hat{u}({\bf k})=\int e^{-i{\bf k}\cdot {\bf x}}{u}({\bf x})\, \frac{d{\bf x}}{(2\pi)^d},\label{sw9}
\end{eqnarray}
we get
\begin{eqnarray}
{\bf F}(1\rightarrow 0,s)=-i m\int {\bf k} \, e^{i{\bf k}\cdot ({\bf r}(s)-{\bf r}_{1}(s))}\hat{u}({\bf k})\, d{\bf k}.\qquad \label{sw10}
\end{eqnarray}
Using the equations of motion (\ref{sw5}), and introducing ${\bf x}={\bf r}-{\bf r}_{1}$ and ${\bf w}={\bf v}-{\bf v}_{1}$, we obtain
\begin{eqnarray}
{\bf F}(1\rightarrow 0,t-\tau)
=-i m\int {\bf k} \, e^{i{\bf k}\cdot ({\bf x}-{\bf w}\tau)}\hat{u}({\bf k})\, d{\bf k}.\label{sw11}
\end{eqnarray}
Therefore
\begin{eqnarray}
K^{\mu\nu}=-m^2 \int_{0}^{+\infty}d\tau\int d{\bf x} \int d{\bf k} \int d{\bf k}' k^{\mu}k^{'\nu} e^{i({\bf k}+{\bf k}')\cdot {\bf x}} e^{-i{\bf k}'\cdot {\bf w}\tau}\hat{u}({\bf k})\hat{u}({\bf k}').\label{sw12}
\end{eqnarray}
Using the identity
\begin{eqnarray}
\delta({\bf x})=\int e^{i{\bf k}\cdot {\bf x}}\, \frac{d{\bf k}}{(2\pi)^d}, \label{sw13}
\end{eqnarray}
and integrating on ${\bf x}$, then on ${\bf k}'$, we find that
\begin{eqnarray}
K^{\mu\nu}=(2\pi)^d m^2  \int_{0}^{+\infty}d\tau \int d{\bf k} \,  k^{\mu}k^{\nu}  e^{i{\bf k}\cdot {\bf w}\tau}\hat{u}({\bf k})\hat{u}(-{\bf k}).\label{sw14}
\end{eqnarray}
Since $u({\bf x})$ is real, we have $\hat{u}(-{\bf k})=\hat{u}({\bf k})^{*}$. Furthermore, since $u(-{\bf x})=u({\bf x})$, we have $\hat{u}({\bf k})^{*}=\hat{u}({\bf k})$. Therefore, $\hat{u}({\bf k})$ is real and $\hat{u}(-{\bf k})=\hat{u}({\bf k})$. Therefore,
\begin{eqnarray}
K^{\mu\nu}=(2\pi)^d m^2  \int_{0}^{+\infty}d\tau \int d{\bf k} \,  k^{\mu}k^{\nu}  e^{i{\bf k}\cdot {\bf w}\tau}\hat{u}({\bf k})^2.\label{sw15}
\end{eqnarray}
Making the transformation $\tau\rightarrow -\tau$, then ${\bf k}\rightarrow -{\bf k}$, and adding the resulting expression to Eq. (\ref{sw15}), we get
\begin{eqnarray}
K^{\mu\nu}=\frac{1}{2}(2\pi)^d m^2  \int_{-\infty}^{+\infty}d\tau \int d{\bf k}  \, k^{\mu}k^{\nu}  e^{i{\bf k}\cdot {\bf w}\tau}\hat{u}({\bf k})^2.\label{sw16}
\end{eqnarray}
Using the identity (\ref{sw13}), we finally obtain
\begin{eqnarray}
K^{\mu\nu}=\pi (2\pi)^d m^2 \int d{\bf k} \,  k^{\mu}k^{\nu} \delta({\bf k}\cdot {\bf w}) \hat{u}({\bf k})^2.\label{sw17}
\end{eqnarray}
Therefore, the kinetic equation takes the form
\begin{equation}
\label{sw18} {\partial f\over\partial t}=\pi (2\pi)^d m{\partial\over\partial
v^{\mu}}\int  k^{\mu}k^{\nu} \delta({\bf k}\cdot {\bf w})  \hat{u}({\bf k})^2 \biggl (f_{1}{\partial
f\over\partial v^{\nu}}-f{\partial
f_{1}\over\partial v_{1}^{\nu}}\biggr )\, d{\bf v}_{1}d{\bf k},
\end{equation}
where we have noted $f=f({\bf v},t)$ and $f_1=f({\bf v}_1,t)$. The calculation of $K^{\mu\nu}$ is detailed in \ref{sec_Kmunu}. Finally, the kinetic equation can be written
\begin{equation}
\label{sw19} {\partial f\over\partial t}=K_d {\partial\over\partial
v^{\mu}}\int \frac{w^2\delta^{\mu\nu}-w^{\mu}w^{\nu}}{w^3} \biggl (f_{1}{\partial
f\over\partial v^{\nu}}-f{\partial
f_{1}\over\partial v_{1}^{\nu}}\biggr )\, d{\bf v}_{1},
\end{equation}
with
\begin{eqnarray}
K_3=8 \pi^5  m \int_{0}^{+\infty} k^3  \hat{u}(k)^2 \, dk, \qquad K_2=8 \pi^3  m \int_{0}^{+\infty} k^2  \hat{u}(k)^2 \, dk.
\label{sw20}
\end{eqnarray}
This is the original form given by Landau \cite{landau} for the Coulombian interaction (in $d=3$). The Landau equation is derived here in a very different manner. This equation also applies to stellar systems if we make a {\it local approximation}. In that case, the distribution functions $f=f({\bf v},t)$ and $f_1=f({\bf v}_1,t)$ must be replaced by $f=f({\bf r},{\bf v},t)$ and $f_1=f({\bf r},{\bf v}_1,t)$ and we must restore the  advective (Vlasov) term in the l.h.s. where the gravitational field is determined by the Poisson equation. Using $(2\pi)^3\hat{u}({\bf k})=-4\pi G/k^2$, this yields the Vlasov-Landau-Poisson equation
\begin{eqnarray}
\label{sw19b} \frac{\partial f}{\partial t}+{\bf v}\cdot {\partial f\over\partial {\bf r}}-\nabla\Phi \cdot {\partial f\over \partial {\bf v}}=K {\partial\over\partial
v^{\mu}}\int \frac{w^2\delta^{\mu\nu}-w^{\mu}w^{\nu}}{w^3} \biggl (f_{1}{\partial
f\over\partial v^{\nu}}-f{\partial
f_{1}\over\partial v_{1}^{\nu}}\biggr )\, d{\bf v}_{1},\nonumber\\
\end{eqnarray}
\begin{equation}
\label{sw19c} \Delta\Phi=4\pi G\int f({\bf r},{\bf v},t)\, d{\bf v},
\end{equation}
where $K=2\pi m G^2\ln\Lambda$ and $\ln \Lambda=\int_0^{+\infty}\frac{dk}{k}$ is the usual Coulomb logarithm that has to be regularized with appropriate cut-offs. It scales like $\ln\Lambda\sim\ln N$  (see, e.g., \cite{paper3}).

The kinetic equation generalizing the Landau Eq. (\ref{sw20}) by
taking into account collective effects is the Lenard-Balescu equation
\cite{lenard,balescu}:
\begin{eqnarray}
\frac{\partial f}{\partial t}=\pi (2\pi)^d m
\frac{\partial}{\partial {v}^{\mu}} \int
k^{\mu}k^{\nu}\delta ({\bf k}\cdot {\bf w}) \frac{\hat{u}({\bf k})^2}{|\epsilon({\bf k},{\bf k}\cdot {\bf
v})|^{2}}  \left (f_1\frac{\partial
f}{\partial v^{\nu}}-f\frac{\partial f_1}{\partial
v_1^{\nu}}\right )\, d{\bf v}_1 d{\bf k},\nonumber\\ \label{sw21}
\end{eqnarray}
where $\epsilon({\bf k},\omega)$ is the dielectric function
\begin{eqnarray}
\epsilon({\bf k},\omega)=1+(2\pi)^{d}\hat{u}({\bf k})\int \frac{{\bf k}\cdot \frac{\partial f}{\partial {\bf v}}}{\omega-{\bf k}\cdot {\bf v}}\, d{\bf v}. \label{sW22}
\end{eqnarray}
The Landau equation is recovered by taking $|\epsilon({\bf k},{\bf
k}\cdot {\bf v})|^{2}=1$. In plasma physics, it has been shown that
collective effects regularize the logarithmic divergence at large
scales that appears in the Landau equation. The Landau and the
Lenard-Balescu equations conserve mass $M=\int f\, d{\bf r}d{\bf v}$
and energy (reducing to the kinetic energy $E=\int f\frac{v^2}{2}\,
d{\bf r}d{\bf v}$ for a spatially homogeneous system) and
monotonically increase the Boltzmann entropy $S=-\int \frac{f}{m}\ln
\frac{f}{m}\, d{\bf r}d{\bf v}$ ($H$-theorem). The collisional
evolution is due to a condition of resonance between the orbits of the
particles. For spatially homogeneous systems, the condition of
resonance encapsulated in the $\delta$-function corresponds to ${\bf
k}\cdot {\bf v}={\bf k}\cdot {\bf v}_{1}$ with ${\bf v}\neq {\bf
v}_{1}$. For $d>1$, the only stationary solution of the Landau and
Lenard-Balescu equations is the Maxwell distribution $f=Ae^{-\beta m \frac{v^2}{2}}$ and these
equations relax towards the Maxwell distribution for $t\rightarrow
+\infty$.  Since the collision term is valid at the order $O(1/N)$,
the relaxation time scales like
\begin{eqnarray}
t_{R}\sim Nt_D, \qquad (d>1).
 \label{sw23}
\end{eqnarray}
This scaling, predicted in \cite{paper2}, has been confirmed for a two-dimensional plasma
in \cite{benedetti,epjb}.
For the Newtonian potential in $d=3$ there is a well-known logarithmic
divergence at large-scales in the Landau equation (\ref{sw19b}). This
divergence is due to the assumption of spatial homogeneity and to the
linear trajectory approximation, and is not present in the generalized
Landau equation (\ref{sw2}). When this divergence is properly
regularized, the relaxation time is found to be $t_{R}\sim (N/\ln
N)t_D$ (see, e.g., \cite{paper3}).  This is the Chandrasekhar relaxation time.
For one-dimensional systems, like
the HMF model, the kinetic equation (\ref{sw21}) reduces to
\begin{eqnarray}
\frac{\partial f}{\partial t}=2\pi^2 m
\frac{\partial}{\partial {v}} \int d{v}_1 d{k}
|k| \frac{\hat{u}(k)^2}{|\epsilon(k,k v)|^{2}} \delta(v-v_{1}) \left
(f_1\frac{\partial f}{\partial v}-f\frac{\partial
f_1}{\partial v_1}\right )=0, \label{sw24}
\end{eqnarray}
where we have used the identity $\delta(\lambda
x)=\frac{1}{|\lambda|}\delta(x)$. Therefore, the collision term
$C_{N}[f]$ vanishes at the order $1/N$ because there is no
resonance. The kinetic equation reduces to $\partial f/\partial t=0$
so that the distribution function does not evolve at all on a
timescale $\sim Nt_{D}$. This result is known for a long time in
plasma physics and was discovered by Eldridge \& Feix \cite{feix} (see
also \cite{kp}) for a one-dimensional plasma: {\it ``To first order in
the small parameter $g=(nD)^{-1}$, the detailed balance between drag
and diffusion is valid not only at thermal equilibrium but for any
stable function $f(v)$. So the Maxwellization is, at least, a
second-order effect in $g$ and consequently a very slow process''.}
This result has been rediscovered recently for the HMF model by
Bouchet \& Dauxois \cite{bd} and Chavanis {\it et al.} \cite{cvb}.
This implies that, for one-dimensional homogeneous systems, the
relaxation time towards statistical equilibrium is larger than
$Nt_D$. We therefore expect that
\begin{eqnarray}
t_{R}> N t_D, \qquad (d=1).
 \label{sw25}
\end{eqnarray}
For the HMF model, it has been found numerically that $t_R\sim N^{1.7}t_D$ \cite{yamaguchi} or $t_R\sim e^N t_D$ \cite{campa}, depending on the type of initial conditions. For 1D inhomogeneous distributions, the relaxation time can be reduced and approach the natural scaling $Nt_D$ associated to the generalized kinetic equation (\ref{sw2}) (see discussion in Sec. \ref{conclusion}).

\subsection{Test particle in a thermal bath: the Fokker-Planck equation}
\label{sec_st}

We now consider a ``test'' particle (tagged particle) evolving in a
steady distribution of ``field'' particles. The test particle has
a stochastic motion. We assume the system to be
spatially homogeneous.  Let us call $P({\bf v},t)$ the probability
density of finding the test particle with velocity ${\bf v}$ at time
$t$. The evolution of $P({\bf v},t)$ can be obtained from the
Lenard-Balescu equation (\ref{sw21}) by considering that the
distribution $f_{1}$ of the field particles is {\it fixed}. Thus, we
replace $f=f({\bf v},t)$ by $P=P({\bf v},t)$ and $f_{1}=f({\bf
v}_{1},t)$ by $f_1=f({\bf v}_{1})$ where $f({\bf v})$ is any stable
stationary solution of the Vlasov equation. This procedure
\cite{paper2} transforms the integro-differential equation
(\ref{sw21}) into the differential equation
\begin{eqnarray}
\label{st1} {\partial P\over\partial t}=\pi (2\pi)^{d}m{\partial\over\partial v^{\mu}}\int k^{\mu}k^{\nu}\delta ({\bf k}\cdot {\bf w}) {\hat{u}({\bf k})^{2}\over |\epsilon({\bf k},{\bf k}\cdot {\bf v})|^{2}}\biggl (f_1 {\partial P \over\partial v^{\nu}}-P{\partial f_1\over\partial {v}_{1}^{\nu}}\biggr )\, d{\bf v}_{1}d{\bf k},\nonumber\\
\end{eqnarray}
where $\epsilon({\bf k},\omega)$ is the dielectric function corresponding to the fixed
distribution function $f({\bf v})$.
Equation (\ref{st1}) can be written in the form of a Fokker-Planck
equation
\begin{equation}
\label{st2}
\frac{\partial  P}{\partial t}={\partial\over\partial v^{\mu}}\biggl (D^{\mu\nu}{\partial P\over\partial v^{\nu}}-P\eta^{\mu}\biggr ),
\end{equation}
involving a diffusion term
\begin{equation}
\label{st3} D^{\mu\nu}=\pi (2\pi)^{d}m\int k^{\mu}k^{\nu}\delta ({\bf k}\cdot {\bf w}){\hat{u}({\bf k})^{2}\over |\epsilon({\bf k},{\bf k}\cdot {\bf v})|^{2}} f_1 \, d{\bf v}_{1}d{\bf k},
\end{equation}
and a friction term due to the polarization
\begin{equation}
\eta^{\mu}=F_{pol}^{\mu}=\pi (2\pi)^{d}m\int k^{\mu}k^{\nu}\delta ({\bf k}\cdot {\bf w}) {\hat{u}({\bf k})^{2}\over |\epsilon({\bf k},{\bf k}\cdot {\bf v})|^{2}}{\partial f_1\over\partial {v}_{1}^{\nu}} \, d{\bf v}_{1}d{\bf k}.
\label{st4}
\end{equation}
The diffusion term can be directly derived from the Kubo formula, and
the friction force due to the polarization can be directly obtained
from a linear response theory, as shown by Kandrup \cite{kandrup2} and
Chavanis \cite{paper3}. Since the diffusion coefficient depends on the
velocity ${\bf v}$ of the test particle, it is useful to rewrite
Eq. (\ref{st2}) in a form that is fully consistent with the general
Fokker-Planck equation
\begin{equation}
\label{st5} {\partial P\over \partial t}={1\over 2}{\partial^{2}\over\partial v^{\mu}\partial v^{\nu}}\biggl (P{\langle \Delta v^{\mu} \Delta v^{\nu}\rangle\over \Delta t}\biggr )-{\partial\over\partial v^{\mu}}\biggl (P{\langle \Delta v^{\mu}\rangle\over \Delta t}\biggr ),
\end{equation}
with
\begin{equation}
\label{st6}{\langle \Delta v^{\mu} \Delta v^{\nu}\rangle\over 2\Delta t}=D^{\mu\nu}, \qquad {\langle \Delta v^{\mu}\rangle\over \Delta t}={\partial D^{\mu\nu}\over\partial v^{\nu}}+\eta^{\mu}\equiv F_{friction}^{\mu}.
\end{equation}
Substituting Eqs. (\ref{st3}) and (\ref{st4}) into Eq. (\ref{st6}), and using an integration by parts,  we find that the diffusion and friction coefficients are given by
\begin{equation}
\label{st7} {\langle \Delta v^{\mu} \Delta v^{\nu}\rangle\over 2\Delta t}=\pi (2\pi)^{d}m\int k^{\mu}k^{\nu}\delta ({\bf k}\cdot {\bf w}){\hat{u}({\bf k})^{2}\over |\epsilon({\bf k},{\bf k}\cdot {\bf v})|^{2}} f_1 \, d{\bf v}_{1}d{\bf k},
\end{equation}
\begin{equation}
{\langle \Delta v^{\mu}\rangle\over \Delta t}=\pi (2\pi)^{d}m\int k^{\mu}k^{\nu} f_1 \left (\frac{\partial}{\partial {v}^{\nu}}-\frac{\partial}{\partial v_1^\nu}\right ) \delta ({\bf k}\cdot {\bf w}) {\hat{u}({\bf k})^{2}\over |\epsilon({\bf k},{\bf k}\cdot {\bf v})|^{2}}\, d{\bf v}_{1}d{\bf k}.
\label{st8}
\end{equation}
The two forms (\ref{st2}) and (\ref{st5}) of the Fokker-Planck equation have their own
interest. The expression (\ref{st5}) where the diffusion coefficient
is placed {after} the two derivatives $\partial^{2}(DP)$ involves the
total friction force $F_{friction}^{\mu}=\langle \Delta
v^{\mu}\rangle/\Delta t$ and the expression (\ref{st2}) where the
diffusion coefficient is placed {between} the derivatives $\partial
D\partial P$ isolates the part of the friction $\eta^{\mu}=
F^{\mu}_{pol}$ due to the polarization \cite{paper4}.

In stellar dynamics, the coefficients of diffusion and friction were
first obtained by Chandrasekhar \cite{chandra,chandra1} by considering
the mean effect of a succession of binary encounters. They were also
computed by Cohen {\it et al.} \cite{cohen} and Rosenbluth {\it et
al.} \cite{rosen} in a plasma physics context by neglecting collective
effects. This leads to expressions that are consistent with
Eqs. (\ref{st7}) and (\ref{st8}) with $|\epsilon({\bf k},{\bf k}\cdot
{\bf v})|^{2}=1$ (see, e.g., \cite{epjb} for a discussion of this
link). These expressions can also be obtained directly from the
Hamiltonian equations of motion by making a systematic expansion of
the trajectory of the particles in powers of $1/N$ in the limit
$N\rightarrow +\infty$ as shown in Appendix A of \cite{paper4}. On the
other hand, the expressions (\ref{st7}) and (\ref{st8}) for the
diffusion and the friction, properly taking into account collective
effects, were first derived by Hubbard \cite{hubbard} in plasma
physics by directly evaluating $\langle \Delta v^{\mu} \Delta
v^{\nu}\rangle$ and $\langle \Delta v^{\mu}\rangle$. Collective
effects encapsulated in the dielectric function remove the logarithmic
divergence that otherwise appears at large scales. The work of Hubbard
\cite{hubbard} was done independently from the works of Lenard
\cite{lenard} and Balescu \cite{balescu} that were published at about
the same time, but they are clearly connected since the equations
(\ref{st7}) and (\ref{st8}) of Hubbard \cite{hubbard} can be derived
from the Lenard-Balescu equation, and {\it vis et versa}, as explained
above (see also discussion in \cite{paper4}).

Let us now consider particular
cases.

$\bullet$ In $d=1$, the expressions of the diffusion and friction simplify into
\begin{equation}
\label{st9} D(v)=4\pi^{2}m f(v)\int_{0}^{+\infty} {k\hat{u}(k)^{2}\over |\epsilon(k,kv)|^{2}}\, dk,
\end{equation}
\begin{equation}
\label{st10} \eta(v)=4\pi^{2}m f'(v)\int_{0}^{+\infty}  {k\hat{u}(k)^{2}\over |\epsilon(k,kv)|^{2}}\, dk=D(v)\frac{d\ln f}{dv}.
\end{equation}
\begin{equation}
{\langle \Delta v\rangle\over \Delta t}=D'(v)+D(v)\frac{d\ln f}{dv},
\label{st11}
\end{equation}
and the Fokker-Planck
equation (\ref{st1}) reduces to
\begin{equation}
\label{st12}  {\partial P\over\partial t}={\partial\over\partial v}\biggl\lbrack D(v)\biggl ({\partial P\over\partial v}-P {d\ln f\over dv}\biggr )\biggr\rbrack.
\end{equation}
The diffusion and the friction coefficients were derived by Eldridge
\& Feix \cite{feix} for one dimensional plasmas and by Bouchet \&
Dauxois \cite{bd} for the HMF model by directly evaluating $\langle
(\Delta v)^2 \rangle$ and $\langle \Delta v\rangle$ (see
\cite{cdr} for detailed calculations). Alternatively, Chavanis
\cite{paper2} (see also \cite{cvb,epjb}) obtains
Eqs. (\ref{st9})-(\ref{st12}) as a particular case of the general
expressions (\ref{st1})-(\ref{st8}) valid in any dimension of space.

$\bullet$ If we neglect collective effects, the diffusion and friction coefficients reduce to
\begin{equation}
\label{st13} D^{\mu\nu}=\pi (2\pi)^{d}m\int k^{\mu}k^{\nu}\delta ({\bf k}\cdot {\bf w})\hat{u}({\bf k})^{2} f_1 \, d{\bf v}_{1}d{\bf k},
\end{equation}
\begin{equation}
\eta^{\mu}=F_{pol}^{\mu}=\pi (2\pi)^{d}m\int k^{\mu}k^{\nu}\delta ({\bf k}\cdot {\bf w}) {\hat{u}({\bf k})^{2}}{\partial f_1\over\partial {v}_{1}^{\nu}} \, d{\bf v}_{1}d{\bf k},
\label{st4sscoll}
\end{equation}
\begin{equation}
F^{\mu}_{friction}=\pi (2\pi)^{d}m\int k^{\mu}k^{\nu}f_1\left (\frac{\partial}{\partial {v}^{\nu}}-\frac{\partial}{\partial v_1^\nu}\right ) \delta ({\bf k}\cdot {\bf w}) {\hat{u}({\bf k})^{2}}\, d{\bf v}_{1}d{\bf k}.
\label{st14}
\end{equation}
After a series of elementary transformations, we obtain
\begin{eqnarray}
\label{st15}
\frac{\partial D^{\mu\nu}}{\partial v^\nu}&=&\pi (2\pi)^{d}m\int k^{\mu}k^{\nu}\frac{\partial}{\partial v^\nu}\delta ({\bf k}\cdot {\bf w})\hat{u}({\bf k})^{2} f_1 \, d{\bf v}_{1}d{\bf k},\nonumber\\
&=&-\pi (2\pi)^{d}m\int k^{\mu}k^{\nu}\frac{\partial}{\partial v_1^\nu}\delta ({\bf k}\cdot {\bf w})\hat{u}({\bf k})^{2} f_1 \, d{\bf v}_{1}d{\bf k},\nonumber\\
&=&\pi (2\pi)^{d}m\int k^{\mu}k^{\nu}\delta ({\bf k}\cdot {\bf w})\hat{u}({\bf k})^{2} \frac{\partial f_1}{\partial v_1^\nu} \, d{\bf v}_{1}d{\bf k}=\eta^\mu.
\end{eqnarray}
Combining Eq. (\ref{st6}) with Eq. (\ref{st15}), we find that
\begin{equation}
\label{st16}
{\langle \Delta {\bf v}\rangle\over \Delta t}=2{\eta^{\mu}},\qquad  {\rm i.e.}\qquad  {\bf F}_{friction}=2{\bf F}_{pol},
\end{equation}
so that the friction force ${\bf F}_{friction}$ is equal to twice the friction due to the polarization ${\bf F}_{pol}$. This explains the difference of factor $2$ in the calculations of Chandrasekhar \cite{chandra1} and Kandrup \cite{kandrup1} (see the anomaly noted by Kandrup \cite{kandrup1} p. 446).

$\bullet$ For a thermal bath, the field particles have the Maxwell distribution of statistical equilibrium
\begin{equation}
\label{st17} f({\bf v}_1)=\biggl ({\beta m\over 2\pi}\biggr )^{d/2}\rho \ e^{-\beta m {v_1^{2}\over 2}}.
\end{equation}
Substituting
\begin{equation}
\label{st18}
\frac{\partial f_1}{\partial v_1^\nu}=-\beta m f_1 v_1^\nu,
\end{equation}
in Eq. (\ref{st4}), we obtain
\begin{equation}
\eta^{\mu}=-\beta \pi (2\pi)^{d}m^2\int k^{\mu} ({\bf k}\cdot {\bf v}_1)\delta \left\lbrack {\bf k}\cdot ({\bf v}-{\bf v}_1)\right\rbrack {\hat{u}({\bf k})^{2}\over |\epsilon({\bf k},{\bf k}\cdot {\bf v})|^{2}}f_1  \, d{\bf v}_{1}d{\bf k}.
\label{st19}
\end{equation}
Using the $\delta$-function to replace ${\bf k}\cdot {\bf v}_1$ by ${\bf k}\cdot {\bf v}$, we get
\begin{equation}
\eta^{\mu}=-\beta \pi (2\pi)^{d}m^2 v^{\nu}\int k^{\mu}k^{\nu} \delta \left\lbrack {\bf k}\cdot ({\bf v}-{\bf v}_1)\right\rbrack {\hat{u}({\bf k})^{2}\over |\epsilon({\bf k},{\bf k}\cdot {\bf v})|^{2}}f_1  \, d{\bf v}_{1}d{\bf k}.
\label{st19bis}
\end{equation}
Comparing the resulting expression with Eq. (\ref{st3}), we find that
\begin{equation}
\eta^{\mu}=-\beta m D^{\mu\nu} v^\nu,
\label{st20}
\end{equation}
which is the appropriate Einstein relation for our problem. Note that
it is valid for the friction force due to the polarization ${\bf
F}_{pol}$, not for the total friction force ${\bf F}_{friction}$. The Fokker-Planck
equation (\ref{st1}) takes the form
\begin{equation}
\label{st21}
\frac{\partial  P}{\partial t}={\partial\over\partial v^{\mu}}\left\lbrack D^{\mu\nu}({\bf v})\left ({\partial P\over\partial v^{\nu}}+\beta m P v^{\nu}\right )\right\rbrack,
\end{equation}
where the diffusion coefficient is given by Eq. (\ref{st3}) with
Eq. (\ref{st17}). This is similar to the Kramers equation except that
the diffusion coefficient is a tensor and that it depends on the
velocity. Some simplified forms of the diffusion coefficient are given in
\cite{epjb} in different dimensions of space $d$.

\section{Two-dimensional point vortices}
\label{sec_vortex}

\subsection{The evolution of the system as a whole: the kinetic equation}
\label{sec_vw}

We consider a system of point vortices described by the
Hamiltonian equations
\begin{eqnarray}
\gamma\frac{d{x}_i}{dt}=\frac{\partial H}{\partial y_i},\qquad \gamma\frac{d{y}_i}{dt}=-\frac{\partial H}{\partial x_i},\nonumber\\
H=\gamma^2\sum_{i<j}u({\bf r}_i-{\bf r}_j).
\label{vw1}
\end{eqnarray}
The usual interaction corresponds to $u({\bf r}-{\bf
r}')=-\frac{1}{2\pi}\ln|{\bf r}-{\bf r}'|$ but we let $u$ arbitrary in
order to treat more general cases. We consider the proper
thermodynamic limit $N\rightarrow +\infty$ in such a way that $V\sim
E\sim \beta/N\sim 1$ and $\gamma\sim 1/N$ \cite{assise} (for rigorous
mathematical results see \cite{caglioti,k93,es,ca2,kl}). When
collective effects are ignored, the kinetic equation describing the
evolution of the smooth vorticity profile $\omega({\bf r},t)$ at the
order $O(1/N)$ can be written
\begin{eqnarray}
\frac{\partial \omega}{\partial t}+\frac{N-1}{N}\langle {\bf
V}\rangle\cdot {\partial \omega\over \partial {\bf r}}
=\frac{\partial}{\partial {r}^{\mu}}\int_0^{t} d\tau \int d{\bf
r}_{1} {V}^{\mu}(1\rightarrow 0)G(t,t-\tau)\nonumber\\
\times  \left \lbrack {{\cal V}}^{\nu}(1\rightarrow 0)
{\partial\over\partial {r}^{\nu}}+{{\cal V}}^{\nu}(0\rightarrow
1) {\partial\over\partial {r}_{1}^{\nu}}\right
\rbrack\omega({\bf
r},t-\tau)\frac{\omega}{\gamma}({\bf r}_1,t-\tau), \label{vw2}
\end{eqnarray}
where ${\bf V}(1\rightarrow 0)$ is the velocity created by point
vortex $1$ (located in ${\bf r}_1$) on point vortex $0$ (located in
${\bf r}$) and ${\cal V}^\mu(1\rightarrow 0)={\bf V}(1\rightarrow
0)-\langle {\bf V}\rangle/N$ is the fluctuating velocity with respect
to the average velocity $\langle {\bf V}\rangle ({\bf r},t)=-{\bf
z}\times\nabla\psi({\bf r},t)$ at the location of point vortex $0$. On
the other hand, $G(t,t-\tau)$ is the Greenian constructed with the
mean flow. This kinetic equation has been obtained by Chavanis
\cite{kin,kinvortex} from a projection operator formalism, a
BBGKY-like hierarchy and a quasilinear theory. The BBGKY hierarchy is
closed by neglecting the three body correlation function which is of
order $O(1/N^2)$. The structure of this equation bears a clear
physical meaning.  It involves a diffusion term and a drift
term. Furthermore, the coefficients of diffusion and drift are given
by generalized Kubo formulae, i.e. the time integral of the temporal
correlations of the fluctuating velocity. The collision term
$C_N[\omega]$ (r.h.s.) is valid at the order $1/N$ so that it
describes the ``collisional'' evolution of the point vortex gas
(ignoring collective effects) on a timescale of order $Nt_D$. For
$N\rightarrow +\infty$, we obtain the 2D Euler equation in which
collisions (more properly correlations) are neglected. The 2D
Euler-Poisson system can generate a process of violent relaxation
towards a quasi stationary state as described by Miller \cite{miller}
and Robert \& Sommeria \cite{rs}. This is the hydrodynamical
counterpart of the Lynden-Bell \cite{lb} theory of violent relaxation
in stellar dynamics (see, e.g. \cite{csr,houches}, for a description
of this analogy). Equation (\ref{vw2}) is a non Markovian
integro-differential equation.  It can describe the dynamics of point
vortices taking into account delocalizations in space and time
(i.e. non-axisymmetry and memory effects). This kinetic equation is
the vortex analogue of the generalized Landau equation (\ref{sw2}) for
stellar systems. If we make the Markovian approximation and extend the
time integral to $+\infty$, we obtain
\begin{eqnarray}
\frac{\partial \omega}{\partial t}+\frac{N-1}{N}\langle {\bf
V}\rangle\cdot {\partial \omega\over \partial {\bf r}}
=\frac{\partial}{\partial {r}^{\mu}}\int_0^{+\infty} d\tau \int d{\bf
r}_{1} {V}^{\mu}(1\rightarrow 0)G(t,t-\tau)\nonumber\\
\times  \left \lbrack {{\cal V}}^{\nu}(1\rightarrow 0)
{\partial\over\partial {r}^{\nu}}+{{\cal V}}^{\nu}(0\rightarrow
1) {\partial\over\partial {r}_{1}^{\nu}}\right
\rbrack\omega({\bf
r},t)\frac{\omega}{\gamma}({\bf r}_1,t). \label{vw3}
\end{eqnarray}
The Markovian approximation may not be justified in any situation
since it has been shown numerically that point vortices can exhibit
long jumps (L\'evy flights) and correlations (see
Sec. \ref{conclusion} for a more detailed discussion). However, for
$N\rightarrow +\infty$, the Markovian approximation should be
justified because the timescale $\sim Nt_D$ on which $\omega({\bf
r},t)$ changes is long compared to the timescale $\tau_{corr}$ for
which the integrand in Eq. (\ref{vw2}) has significant support. The
Markovian equation (\ref{vw3}) still applies to possibly
non-axisymmetric distributions, which is the norm for 2D flows. Now,
for axisymmetric distributions $\omega(r,t)$, the kinetic equation
reduces to
\begin{eqnarray}
\frac{\partial \omega}{\partial t}=\frac{1}{r}\frac{\partial}{\partial {r}} r \int_0^{+\infty} d\tau \int d{\bf
r}_{1} {V}_r(1\rightarrow 0)G(t,t-\tau)\nonumber\\
\times \left \lbrack {{V}}_r(1\rightarrow 0)
{\partial\over\partial {r}}+{{V}}_{r_1}(0\rightarrow
1) {\partial\over\partial {r}_{1}}\right
\rbrack\omega({r},t)\frac{\omega}{\gamma}({r}_1,t). \label{vw4}
\end{eqnarray}
Using $V_{r_1}(0\rightarrow 1)=-\frac{r}{r_1}V_r(1\rightarrow 0)$ (see
below) and the equation of the unperturbed trajectory
\begin{eqnarray}
r(t-\tau)=r(t)=r,\qquad \theta(t-\tau)=\theta(t)-\Omega(r,t)\tau=\theta-\Omega(r,t)\tau,
\label{vw5}
\end{eqnarray}
corresponding to $\langle {\bf V}\rangle=V(r,t){\bf e}_{\theta}$ with $V(r,t)=\Omega(r,t)r$ (where $\Omega$ is the angular velocity), the kinetic equation can be written
\begin{eqnarray}
\frac{\partial \omega}{\partial t}=\frac{1}{r}\frac{\partial}{\partial {r}}  \int_{0}^{+\infty}  r_1\, dr_1 K \left (
\frac{1}{r}{\partial\over\partial {r}}-\frac{1}{r_1} {\partial\over\partial {r}_{1}}\right )\omega({r},t)\frac{\omega}{\gamma}({r}_1,t),
\label{vw6}
\end{eqnarray}
with
\begin{eqnarray}
K=r^2\int_{0}^{+\infty}d\tau \int_{0}^{2\pi}d\theta_1\,  {V}_r(1\rightarrow 0,t){{V}}_r(1\rightarrow 0,t-\tau).
\label{vw7}
\end{eqnarray}
In consistency with the Markov approximation, we have assumed that the
angular velocity $\Omega(r,t)$ does not change on the timescale
$\tau_{corr}$ on which the velocity correlations have their
support. The velocity created by point vortex $1$ on point vortex $0$
is given by ${\bf V}(1\rightarrow 0)=-\gamma{\bf z}\times\nabla u({\bf
r}-{\bf r}_1)$ leading to
\begin{eqnarray}
{V}_r(1\rightarrow 0)=\frac{\gamma}{r}\frac{\partial u}{\partial\theta}({\bf r}-{\bf r}_1).
\label{vw8}
\end{eqnarray}
The potential of interaction is of the form $u({\bf r}-{\bf r}_1)=u(r,r_1,\theta-\theta_1)$. It depends on $\cos(\theta-\theta_1)$ and is invariant with respect to the interchange of $0$ and $1$. Introducing the Fourier transform with respect to the angles
\begin{eqnarray}
u(r,r_1,\phi)=\sum_m e^{im\phi}\hat{u}_m(r,r_1),\quad \hat{u}_m(r,r_1)=\frac{1}{2\pi}\int_{0}^{2\pi}u(r,r_1,\phi) \cos(m\phi)\, d\phi,\nonumber\\
\label{vw10}
\end{eqnarray}
we get
\begin{eqnarray}
V_r(1\rightarrow 0,s)=i\frac{\gamma}{r(s)}\sum_m m e^{im (\theta(s)-\theta_1(s))}\hat{u}_m(r(s),r_1(s)).
\label{vw11}
\end{eqnarray}
Using the equations of motion (\ref{vw5}), and introducing $\phi=\theta-\theta_1$ and $\Delta\Omega=\Omega(r,t)-\Omega(r_1,t)$, we obtain
\begin{eqnarray}
V_r(1\rightarrow 0,t-\tau)=i\frac{\gamma}{r}\sum_m m e^{im(\phi-\Delta\Omega\tau)}\hat{u}_m(r,r_1).
\label{vw12}
\end{eqnarray}
Therefore
\begin{eqnarray}
K=-\gamma^2\int_{0}^{+\infty} d\tau\int_{0}^{2\pi}d\phi \sum_{mn} m n e^{im\phi}e^{in(\phi-\Delta\Omega\tau)}\hat{u}_m(r,r_1)\hat{u}_n(r,r_1).
\label{vw13}
\end{eqnarray}
Using the identity
\begin{eqnarray}
\delta_{n,0}=\int_{0}^{2\pi} e^{i n \theta}\, \frac{d\theta}{2\pi},
\label{vw14}
\end{eqnarray}
and integrating on $\phi$, then summing on $n$, we find that
\begin{eqnarray}
K=2\pi\gamma^2\int_{0}^{+\infty} d\tau \sum_{m} m^2 e^{im\Delta\Omega\tau}\hat{u}_m(r,r_1)\hat{u}_{-m}(r,r_1).
\label{vw15}
\end{eqnarray}
Since $\hat{u}_{-m}(r,r_1)=\hat{u}_{m}(r,r_1)$, the foregoing equation can be rewritten
\begin{eqnarray}
K=2\pi\gamma^2\int_{0}^{+\infty} d\tau \sum_{m} m^2 e^{im\Delta\Omega\tau}\hat{u}_m(r,r_1)^2.
\label{vw16}
\end{eqnarray}
Making the transformation $\tau\rightarrow -\tau$, then $m\rightarrow -m$, and adding the resulting expression to Eq. (\ref{vw16}), we get
\begin{eqnarray}
K=\pi\gamma^2\int_{-\infty}^{+\infty} d\tau \sum_{m} m^2 e^{im\Delta\Omega\tau}\hat{u}_m(r,r_1)^2.
\label{vw17}
\end{eqnarray}
Using the identity (\ref{sw13}), we finally  obtain
\begin{eqnarray}
K=2\pi^2\gamma^2 \sum_{m} m^2 \delta(m\Delta\Omega) \hat{u}_m(r,r_1)^2.
\label{vw18}
\end{eqnarray}
Therefore, the kinetic equation takes the form
\begin{eqnarray}
\frac{\partial \omega}{\partial t}=2\pi^2\gamma\frac{1}{r}\frac{\partial}{\partial {r}}  \int_{0}^{+\infty}  r_1\, dr_1 \chi(r,r_1)\delta(\Omega-\Omega_1) \left (
\frac{1}{r}\omega_1{\partial\omega\over\partial {r}}-\frac{1}{r_1}\omega {\partial\omega_1\over\partial {r}_{1}}\right ),
\label{vw19}
\end{eqnarray}
with
\begin{eqnarray}
\chi(r,r_1)=\sum_{m=-\infty}^{+\infty} |m|  \hat{u}_m(r,r_1)^2,
\label{vw20}
\end{eqnarray}
where we have used the identity $\delta(\lambda
x)=\frac{1}{|\lambda|}\delta(x)$ and noted $\omega=\omega(r,t)$,
$\omega_1=\omega(r_1,t)$, $\Omega=\Omega(r,t)$ and
$\Omega_1=\Omega(r_1,t)$.  The calculation of $\chi(r,r_1)$ is
detailed in \ref{sec_chi} for the ordinary potential $u({\bf r}-{\bf
r}')=-\frac{1}{2\pi}\ln({\bf r}-{\bf r}')$. Equation (\ref{vw19}) has
been derived by Chavanis \cite{kin}. It is the vortex analogue of the
Landau equation (\ref{sw18}).  A more general kinetic equation, taking
into account collective effects, has been derived by Dubin \& O'Neil
\cite{dn} by a different method. It is the vortex analogue of the
Lenard-Balescu equation (\ref{sw21}). The kinetic equation
(\ref{vw19}) conserves circulation $\Gamma=\int\omega\, d{\bf r}$,
energy $E=\frac{1}{2}\int\omega\psi\, d{\bf r}$ and angular momentum
$L=\int\omega r^2\, d{\bf r}$, and monotonically increases the
Boltzmann entropy $S=-\int \frac{\omega}{\gamma}\ln\frac{\omega}{\gamma}\, d{\bf r}$ ($H$-theorem). The collisional evolution is due to a
condition of resonance between distant orbits of the point
vortices. For axisymmetric systems, the condition of resonance
encapsulated in the $\delta$-function corresponds to
$\Omega(r,t)=\Omega(r_1,t)$ with ${r}\neq r_{1}$. The evolution stops
when the profile of angular velocity becomes monotonic (so that there
is no resonance) even if the system has not reached the statistical
equilibrium state given by the Boltzmann distribution \cite{mj}. This
``kinetic blocking'' has been illustrated  numerically in
\cite{cl}. Indeed, the Boltzmann distribution $\omega=Ae^{-\beta\gamma(\psi+\frac{1}{2}\Omega_L r^2)}$ is not the unique steady
state of the kinetic equation (\ref{vw19}): any vorticity distribution
associated with a monotonic profile of angular velocity is a steady
state of Eq.  (\ref{vw19}). In particular, if the profile of angular
velocity is initially monotonic\footnote{This is the generic case
because the process of violent relaxation usually leads to an
axisymmetric QSS with a monotonic profile of angular velocity.}, the
collision term $C_{N}[\omega]$ vanishes at the order $1/N$ because
there is no resonance \cite{kin}. The kinetic equation reduces to
$\partial
\omega/\partial t=0$ so that the vorticity does not evolve at all on a
timescale $\sim Nt_{D}$.  This implies that, for axisymmetric distributions of
point vortices, the relaxation time towards statistical equilibrium is larger
than $Nt_D$. We therefore expect that
\begin{eqnarray}
t_{R}> N t_D \qquad ({\rm axisymmetric \ flows}).
 \label{vw21}
\end{eqnarray}
For non axisymmetric distributions, the relaxation time can be reduced
and approach the natural scaling $Nt_D$ associated with the generalized kinetic equation (\ref{vw2}) (see discussion in
Sec. \ref{conclusion}). In any case, for $N\gg 1$, the relaxation
towards the Boltzmann distribution is a very slow process. The
possible slow timescale of mixing was pointed out by Onsager in a
letter to Lin \cite{esr}: {\it ``I still have to find out whether the
processes anticipated by these considerations are rapid enough to play
a dominant role in the evolution.''} and it is now confirmed by the
kinetic theory (as far as we know, the process of violent relaxation
\cite{lb,miller,rs} was not foreseen by Onsager who focused on the
ordinary statistical equilibrium state). In fact, the scaling of the
relaxation time with $N$ is still not known for axisymmetric
distributions and it would be interesting to obtain it numerically. It
is not even clear whether the system of point vortices truly relaxes
towards the Boltzmann distribution predicted by statistical mechanics
\cite{mj}. Indeed, as discussed by Khanin \cite{khanin}, the
dynamics may be non-ergodic. This was also a concern of Onsager: {\it ``We inquire
about the ergodic motion of the system,''} Onsager wrote to Lin \cite{esr}.

\subsection{Relaxation of a test vortex in a thermal bath: the Fokker-Planck equation}
\label{sec_tv}

We now consider the relaxation of a ``test'' vortex (tagged particle)
evolving in a steady distribution of ``field'' vortices. The test
vortex has a stochastic motion. Let us call
$P({\bf r},t)$ the probability density of finding the test vortex at
position ${\bf r}$ at time $t$. For axisymmetric distributions, the
evolution of $P({\bf r},t)$ can be obtained from the kinetic equation
(\ref{vw19}) by considering that the distribution of the field
vortices $\omega_1$ is {\it fixed}. Thus, we replace
$\omega=\omega({r},t)$ by $P=P({r},t)$ and
$\omega_{1}=\omega({r}_{1},t)$ by $\omega_1=\omega({r}_{1})$ where
$\omega({r})$ is any stable stationary solution of the 2D Euler
equation. This procedure \cite{kin,cl} transforms the
integro-differential equation (\ref{vw19}) into the differential
equation
\begin{eqnarray}
\frac{\partial P}{\partial t}=2\pi^2 \gamma \frac{1}{r}
\frac{\partial}{\partial r} \int_0^{+\infty} r_1 dr_1 \chi(r,r_1)
\delta(\Omega-\Omega_1) \left (\frac{1}{r}\frac{\partial}{\partial
r}-\frac{1}{r_1}\frac{\partial}{\partial
 r_1}\right )P(r,t)\omega(r_{1}). \nonumber\\
\label{tv1}
\end{eqnarray}
This is the vortex analogue of Eq. (\ref{st1}). Equation (\ref{tv1}) can be written in the form of a Fokker-Planck equation
\begin{equation}
\label{tv2}{\partial P\over\partial t}=\frac{1}{r}{\partial\over\partial r}\biggl\lbrack r\biggl (D{\partial P\over\partial r}-P\eta\biggr )\biggr\rbrack,
\end{equation}
involving a diffusion term
\begin{eqnarray}
D=\frac{2\pi^{2}\gamma}{r^{2}}\int_{0}^{+\infty}r_{1}dr_{1}\chi(r,r_{1})\delta(\Omega-\Omega_{1})\omega({r}_{1}),
\label{tv3}
\end{eqnarray}
and a drift term due to the polarization
\begin{eqnarray}
\eta\equiv
V_{r}^{pol}=\frac{2\pi^{2}\gamma}{r}\int_{0}^{+\infty}dr_{1}\chi(r,r_{1})\delta(\Omega-\Omega_{1})\frac{d\omega}{d r} ({r}_{1}).
\label{tv4}
\end{eqnarray}
The diffusion coefficient can be directly derived  from the Kubo formula
and the drift velocity due to the polarization
can be directly derived from a linear response theory \cite{preR,kin,kinvortex}. Since the diffusion coefficient depends on the position
${r}$ of the test vortex, it is useful to rewrite Eq. (\ref{tv1}) in a
form that is fully consistent with the general Fokker-Planck equation
\begin{equation}
\label{tv5} {\partial P\over\partial t}={1\over
2r}{\partial\over\partial r}\biggl\lbrack r{\partial\over\partial
r}\biggl  ({\langle (\Delta r)^{2}\rangle\over \Delta t}P\biggr
)\biggr\rbrack-{1\over r}{\partial\over\partial r}\biggl (rP{\langle
\Delta r\rangle\over \Delta t}\biggr ),
\end{equation}
with
\begin{equation}
\label{tv6}{\langle (\Delta r)^{2}\rangle\over 2 \Delta t}=D, \qquad
{\langle \Delta r\rangle\over \Delta t}={\partial D\over\partial
r}+\eta\equiv V_r^{drift}.
\end{equation}
Substituting Eqs. (\ref{tv3}) and (\ref{tv4}) in Eq. (\ref{tv6}), and using an integration by parts, we find that the diffusion and drift coefficients are given by
\begin{eqnarray}
{\langle (\Delta r)^{2}\rangle\over 2 \Delta t}=\frac{2\pi^{2}\gamma}{r^{2}}\int_{0}^{+\infty}r_{1}dr_{1}\chi(r,r_{1})\delta(\Omega-\Omega_{1})\omega({r}_{1}),
\label{tv7}
\end{eqnarray}
\begin{eqnarray}
{\langle (\Delta r)\rangle\over \Delta t}=2\pi^{2}\gamma\int_{0}^{+\infty} r r_{1}dr_{1}\omega_1 \left (\frac{1}{r}\frac{\partial}{\partial r}-\frac{1}{r_1}\frac{\partial}{\partial r_1}\right )\chi(r,r_{1})\delta(\Omega-\Omega_{1})\frac{1}{r^2}.
\label{tv8}
\end{eqnarray}
The two forms (\ref{tv2}) and (\ref{tv5}) of the Fokker-Planck equation are interesting to consider as they involve respectively
the drift due to the polarization $V_r^{pol}$ or the total drift
$V_r^{drift}$ \cite{kinvortex}. Expressions (\ref{tv7}) and (\ref{tv8}) for
the diffusion coefficient and the drift term can  be obtained
directly from the Hamiltonian equations, by making a systematic
expansion of the trajectories of the point vortices in powers of $1/N$
in the limit $N\rightarrow +\infty$ as shown in Appendix C of \cite{cl}.

Let us now consider particular cases.

$\bullet$ If the profile of angular velocity of the field vortices $\Omega(r)$
is monotonic, using the identity
$\delta(\Omega-\Omega_{1})=\delta(r-r_1)/|\Omega'(r)|$, we find that the expressions of the diffusion and drift simplify into
\begin{eqnarray}
D(r)={2\pi^2\gamma}
\frac{\chi(r,r)}{|\Sigma(r)|}\omega(r),\label{tv9}
\end{eqnarray}
\begin{eqnarray}
\eta={2\pi^2\gamma}
\frac{\chi(r,r)}{|\Sigma(r)|}\omega'(r)=D(r)\frac{d\ln\omega}{dr},\label{tv10}
\end{eqnarray}
\begin{equation}
\label{tv11}
{\langle \Delta r\rangle\over \Delta t}=D'(r)+D(r)\frac{d\ln\omega}{dr},
\end{equation}
where $\Sigma(r)=r\Omega'(r)$ is the local shear. The
Fokker-Planck equation (\ref{tv1}) can then be written
\begin{equation}
\label{tv12}{\partial P\over\partial
t}=\frac{1}{r}{\partial\over\partial r}\biggl\lbrack r D(r)\biggl
({\partial P\over\partial r}-P\frac{d\ln\omega}{dr}\biggr
)\biggr\rbrack,
\end{equation}
with a diffusion coefficient given by Eq. (\ref{tv9}). Equation
(\ref{tv12}) is the vortex analogue of Eq. (\ref{st12}). The
expressions (\ref{tv9}) and (\ref{tv10}) of the diffusion coefficient
and of the drift, {\it that are inversely proportional to the shear},
were first derived for a thermal bath in \cite{preR} and for an
arbitrary distribution of the field vortices in \cite{kin} (see remark
before Eq. (123)). For the usual potential of interaction (see \ref{sec_chi}), we have
$\chi(r,r)=\frac{1}{8\pi^2}\ln\Lambda$ where
$\ln\Lambda=\sum_{m=1}^{+\infty}\frac{1}{m}$ is a Coulombian logarithm that has
to be regularized appropriately. It scales like $\ln\Lambda\sim
\frac{1}{2}\ln N$ (see, e.g. \cite{cl}).

$\bullet$ For a thermal bath, the field vortices have the Boltzmann distribution of statistical equilibrium
\begin{eqnarray}
\omega({r}_1)=A\gamma e^{-\beta \gamma \psi'(r_1)}, \label{tv13}
\end{eqnarray}
where $\psi'(r_1)=\psi(r_1)+\frac{\Omega_{L}}{2}r_1^{2}$ is the relative stream function taking into account the invariance by rotation \cite{cl}. We have
\begin{eqnarray}
\frac{d\omega_1}{dr_1}=-\beta\gamma
\omega(r_1)\frac{d\psi'_1}{dr_1}=\beta\gamma\omega(r_1) (\Omega(r_1)-\Omega_{L})r_1,\label{tv14}
\end{eqnarray}
where we have used $\Omega(r)=V(r)/r=-(1/r)d\psi/d r$.
Substituting this relation in Eq. (\ref{tv4}), we obtain
\begin{eqnarray}
\eta=\beta\frac{2\pi^{2}\gamma^2}{r}\int_{0}^{+\infty}dr_{1}
\chi(r,r_{1})\delta(\Omega-\Omega_{1})\omega_1(\Omega_1-\Omega_L)r_1.\label{tv15}
\end{eqnarray}
Using the $\delta$-function to replace $\Omega_1$ by $\Omega$, then using
$\Omega(r)-\Omega_{L}=-(1/r){d\psi'}/{dr}$,  and comparing the resulting
expression with Eq. (\ref{tv3}), we finally find that
\begin{eqnarray}
\eta=-D\beta\gamma \frac{d\psi'}{dr}. \label{tv16}
\end{eqnarray}
The drift is perpendicular to the relative mean field velocity
$\langle {\bf V}'\rangle=-({d\psi'}/{dr}){\bf e}_{\theta}$ and the
drift coefficient satisfies an Einstein relation $\xi=D\beta\gamma$
\cite{preR}.  We stress that the Einstein relation is valid for the
drift $V_{r}^{pol}=\eta$ due to the polarization \cite{kinvortex}, not for
the total drift $V_r^{drift}$.  For a thermal bath, using
Eq. (\ref{tv16}), the Fokker-Planck equation (\ref{tv2}) takes the
form
\begin{equation}
\label{tv17}{\partial P\over\partial
t}=\frac{1}{r}{\partial\over\partial r}\biggl\lbrack r D(r)\biggl
({\partial P\over\partial r}+\beta\gamma P\frac{d\psi'}{dr}\biggr
)\biggr\rbrack,
\end{equation}
where $D(r)$ is given by Eq. (\ref{tv3}) with Eq. (\ref{tv13}). Of
course, if the profile of angular velocity of the Boltzmann
distribution is monotonic, we find that Eq. (\ref{tv10}) with
Eq. (\ref{tv13}) returns Eq. (\ref{tv16}) with a diffusion coefficient
given by Eq. (\ref{tv9}) with Eq. (\ref{tv13}). Note that the
systematic drift $ {\bf V}_{pol}=-D\beta\gamma\nabla\psi'$ of a point
vortex \cite{preR} is the counterpart of the dynamical friction
${\bf F}_{pol}=-D_{\|}\beta m {\bf v}$ of a star
\cite{chandra1} and the Fokker-Planck equation (\ref{tv17}) is the
counterpart of the Fokker-Planck equation (\ref{st21}).

We have presented the results for axisymmetric distributions, but
similar results can be obtained for unidirectional flows
\cite{preR,kin}.

\section{Kinetic equation with angle-action variables}
\label{sec_angleaction}

\subsection{Evolution of the system as a whole: the kinetic equation} \label{sec_ses}

The kinetic equation (\ref{sw18}) derived in Sec. \ref{sec_stellar} is
valid for spatially homogeneous systems. A manner to describe
spatially inhomogeneous systems is to use angle-action variables
$(\phi,{J})$. A kinetic
equation for $f(J,t)$ has
been derived by Chavanis \cite{action} by using two different
strategies: (i) by incorporating in a Fokker-Planck equation the
expressions of the coefficients of diffusion and friction obtained by
Valageas \cite{valageas} or, (ii) by starting from the Klimontovich equation,
using a quasilinear theory, and solving the equations with Laplace-Fourier transforms.  However, the
second derivation presents some shortcomings that were explicitly discussed in \cite{action}.
 In particular, a
factorization hypothesis is used without precise justification.  Here,
we show that, when collective effects are neglected, the kinetic
equation in angle-action variables can be derived from the general
equation
\begin{eqnarray}
\frac{\partial f}{\partial t}=\frac{\partial}{\partial J}\int_{0}^{+\infty}d\tau\int d\phi_1 dJ_1 F(1\rightarrow 0)G(t,t-\tau)\nonumber\\
\times \left\lbrack F(1\rightarrow 0)\frac{\partial}{\partial J}+F(0\rightarrow 1)\frac{\partial}{\partial J_1}\right \rbrack f(J,t)\frac{f}{m}(J_1,t),
\label{ses1}
\end{eqnarray}
with
\begin{eqnarray}
F(1\rightarrow 0)=-m\frac{\partial u}{\partial \phi}(x(J,\phi)-x(J_1,\phi_1)).
\label{ses2}
\end{eqnarray}
Not only the present approach solves the problems of our former approach \cite{action}, but it
also shows a nice unity with the kinetic theories developed in
Secs. \ref{sec_stellar} and \ref{sec_vortex}. Indeed, the kinetic equation (\ref{ses1}) is the analogue of the kinetic equations (\ref{sw3}) and (\ref{vw3}). For simplicity, we consider one-dimensional systems but we shall indicate at the end how our results can be generalized in $d$ dimensions. Introducing the Fourier transform of the potential with respect to the angles
\begin{eqnarray}
u(x(J,\phi)-x(J_1,\phi_1))=\sum_{n,n_1}A_{n,n_1}(J,J_1)e^{i(n\phi-n_1\phi_1)},
\label{ses3}
\end{eqnarray}
where
\begin{eqnarray}
A_{n,n_1}(J,J_1)=\frac{1}{(2\pi)^2}\int u(x(J,\phi)-x(J_1,\phi_1))e^{-i(n\phi-n_1\phi_1)}\, d\phi d\phi_1,
\label{ses4}
\end{eqnarray}
we get
\begin{eqnarray}
F(1\rightarrow 0)=-im \sum_{n,n_1}A_{n,n_1}(J,J_1)n e^{i(n\phi-n_1\phi_1)}.
\label{ses5}
\end{eqnarray}
Substituting this expression in the kinetic equation (\ref{ses1}), we obtain
\begin{eqnarray}
\frac{\partial f}{\partial t}=-m^2\frac{\partial}{\partial J}\int_{0}^{+\infty}d\tau\int d\phi_1 dJ_1 \sum_{n,n_1} \sum_{l,l_1} A_{n,n_1}(J,J_1) n e^{i(n\phi-n_1\phi_1)} G(t,t-\tau)\nonumber\\
\times \left\lbrack A_{l,l_1}(J,J_1) l e^{i(l\phi-l_1\phi_1)}\frac{\partial}{\partial J}+A_{l_1,l}(J_1,J) l_1 e^{i(l_1\phi_1-l\phi)}\frac{\partial}{\partial J_1}\right \rbrack f(J,t)\frac{f}{m}(J_1,t).\nonumber\\
\label{ses6}
\end{eqnarray}
With angle-action variables, the equations of motion take a very simple form
\begin{eqnarray}
J(t-\tau)=J(t)=J,\qquad \phi(t-\tau)=\phi(t)-\Omega(J,t)\tau=\phi-\Omega(J,t)\tau,
\label{ses7}
\end{eqnarray}
where $2\pi\Omega^{-1}(J,t)$ is the period of the orbit with action $J$. Substituting these relations in Eq. (\ref{ses6}) and making the transformations $l\rightarrow -l$ and $l_1\rightarrow -l_1$ in the second integral (friction term), we obtain successively
\begin{eqnarray}
\frac{\partial f}{\partial t}=-m^2\frac{\partial}{\partial J}\int_{0}^{+\infty}d\tau\int d\phi_1 dJ_1 \sum_{n,n_1} \sum_{l,l_1} A_{n,n_1}(J,J_1) n e^{i(n\phi-n_1\phi_1)} \nonumber\\
\times e^{i(l\phi(t-\tau)-l_1\phi_1(t-\tau))} \left\lbrack A_{l,l_1}(J,J_1) l \frac{\partial}{\partial J}-A_{-l_1,-l}(J_1,J) l_1 \frac{\partial}{\partial J_1}\right \rbrack f(J,t)\frac{f}{m}(J_1,t),\nonumber\\
\label{ses8}
\end{eqnarray}
and
\begin{eqnarray}
\frac{\partial f}{\partial t}=-m^2\frac{\partial}{\partial J}\int_{0}^{+\infty}d\tau\int d\phi_1 dJ_1 \sum_{n,n_1} \sum_{l,l_1} A_{n,n_1}(J,J_1) n e^{i(n\phi-n_1\phi_1)}e^{i(l\phi-l_1\phi_1)}\nonumber\\
e^{-i(l\Omega(J,t)-l_1\Omega(J_1,t))\tau} \left\lbrack A_{l,l_1}(J,J_1) l \frac{\partial}{\partial J}-A_{-l_1,-l}(J_1,J) l_1 \frac{\partial}{\partial J_1}\right \rbrack f(J,t)\frac{f}{m}(J_1,t).\nonumber\\
\label{ses9}
\end{eqnarray}
It is easy to establish that
\begin{eqnarray}
A_{n_1,n}(J_1,J)=A_{-n,-n_1}(J,J_1)=A_{n,n_1}(J,J_1)^*.
\label{ses10}
\end{eqnarray}
Therefore, the kinetic equation can be rewritten
\begin{eqnarray}
\frac{\partial f}{\partial t}=-m^2\frac{\partial}{\partial J}\int_{0}^{+\infty}d\tau\int d\phi_1 dJ_1 \sum_{n,n_1} \sum_{l,l_1} A_{n,n_1}(J,J_1) n e^{i(n+l)\phi} e^{-i(n_1+l_1)\phi_1}\nonumber\\
\times e^{-i(l\Omega(J,t)-l_1\Omega(J_1,t))\tau}A_{l,l_1}(J,J_1)
 \left (l \frac{\partial}{\partial J}- l_1 \frac{\partial}{\partial J_1}\right ) f(J,t)\frac{f}{m}(J_1,t).
\label{ses11}
\end{eqnarray}
Integrating on $\phi_1$  and using the fact that average distribution does not depend on $\phi$, we get
\begin{eqnarray}
\frac{\partial f}{\partial t}=2\pi m^2\frac{\partial}{\partial J}\int_{0}^{+\infty}d\tau\int dJ_1 \sum_{n,n_1}  |A_{n,n_1}(J,J_1)|^2  n  \nonumber\\
\times e^{i(n\Omega(J,t)-n_1\Omega(J_1,t))\tau}\left (n \frac{\partial}{\partial J}-n_1 \frac{\partial}{\partial J_1}\right ) f(J,t)\frac{f}{m}(J_1,t).
\label{ses12}
\end{eqnarray}
Making the transformation $\tau\rightarrow -\tau$, then $(n,n_1)\rightarrow (-n,-n_1)$, and adding the resulting expression to Eq. (\ref{ses12}), we obtain
\begin{eqnarray}
\frac{\partial f}{\partial t}=\pi m^2\frac{\partial}{\partial J}\int_{-\infty}^{+\infty}d\tau\int dJ_1 \sum_{n,n_1}  |A_{n,n_1}(J,J_1)|^2  n\nonumber\\
\times e^{i(n\Omega(J,t)-n_1\Omega(J_1,t))\tau}
 \left (n \frac{\partial}{\partial J}-n_1 \frac{\partial}{\partial J_1}\right ) f(J,t)\frac{f}{m}(J_1,t).
\label{ses13}
\end{eqnarray}
Using the identity (\ref{sw13}), we finally obtain the kinetic equation
\begin{eqnarray}
\frac{\partial f}{\partial t}=2\pi^2 m\frac{\partial}{\partial J}\int dJ_1 \sum_{n,n_1}  |A_{n,n_1}(J,J_1)|^2  n \nonumber\\
\times\delta(n\Omega(J,t)-n_1\Omega(J_1,t))
 \left (n f_1\frac{\partial f}{\partial J}-n_1 f\frac{\partial f_1}{\partial J_1}\right ),
\label{ses14}
\end{eqnarray}
which returns the result obtained in \cite{action}.  However, it is
derived here in a more satisfactory manner. In $d$ dimensions, $n$,
$J$, $\Omega$ are replaced by vectors and the coefficient $2\pi^2$ is
replaced by $\pi(2\pi)^d$.

The kinetic equation (\ref{ses14}) conserves mass and energy and
monotonically increases the Boltzmann entropy ($H$-theorem)
\cite{action}. The collisional evolution is due to a condition of
resonance encapsulated in the $\delta$-function. The evolution stops
when this condition of resonance cannot be satisfied anymore even if
the system has not reached the statistical equilibrium state given by
the Boltzmann distribution. Indeed, the Boltzmann distribution is not
the only steady state of Eq. (\ref{ses14}): any distribution
satisfying $n\Omega(J,t)\neq n_1\Omega(J_1,t)$ for any couples
$(n,J)\neq (n_1,J_1)$ is a steady state.  However, the important point
that we want to make here (it will be discussed further in
Sec. \ref{conclusion}) is that, for inhomogeneous systems, there
generically exists much more resonances than for homogeneous
systems. In particular, for one dimensional homogeneous systems, there
is no resonance at all (see Sec. \ref{sec_stellar}). We therefore
expect that the relaxation time will be reduced for inhomogeneous 1D
systems as compared to homogeneous 1D systems. In fact, we cannot
conclude that Eq. (\ref{ses14}) tends to the Boltzmann distribution
(since it is not the only steady state) but since entropy increases
while energy and mass are conserved the system is expected to {\it
approach} the Boltzmann distribution on the natural timescale $N t_D$
on which Eq. (\ref{ses14}) is valid, provided that there are enough
resonances. Indeed, due to ``collisions'' and resonances, the system
becomes ``more mixed''. These arguments will be further discussed in
Sec. \ref{conclusion} in the light of existing numerical results.

\subsection{Relaxation of a particle in a thermal bath: the Fokker-Planck equation} \label{sec_ta}

Implementing a test particle approach like in Secs. \ref{sec_st} and \ref{sec_tv}, we find that the equation for $P(J,t)$ is
\begin{eqnarray}
\frac{\partial P}{\partial t}=2\pi^2 m\frac{\partial}{\partial J}\int dJ_1 \sum_{n,n_1}  |A_{n,n_1}(J,J_1)|^2  n \nonumber\\
\times\delta(n\Omega(J)-n_1\Omega(J_1))
 \left (n f_1\frac{\partial P}{\partial J}-n_1 P\frac{d f_1}{d J_1}\right ).
\label{ta1}
\end{eqnarray}
Equation (\ref{ta1}) can be written in the form of a Fokker-Planck equation
\begin{equation}
\label{ta2}{\partial P\over\partial t}={\partial\over\partial J}\biggl (D{\partial P\over\partial J}-P\eta\biggr ),
\end{equation}
involving a diffusion term
\begin{eqnarray}
D=2\pi^2 m\int dJ_1 \sum_{n,n_1}  |A_{n,n_1}(J,J_1)|^2  n^2 \delta(n\Omega(J)-n_1\Omega(J_1))
 f_1,
\label{ta3}
\end{eqnarray}
and a friction term due to polarization
\begin{eqnarray}
\eta\equiv
F^{pol}=2\pi^2 m\int dJ_1 \sum_{n,n_1}  |A_{n,n_1}(J,J_1)|^2  n \delta(n\Omega(J)-n_1\Omega(J_1))
 n_1 \frac{df_1}{dJ_1}.\nonumber\\
\label{ta4}
\end{eqnarray}
The ordinary Fokker-Planck equation is
\begin{equation}
\label{ta5} {\partial P\over\partial t}={1\over
2}{\partial^2\over\partial J^2}\biggl  ({\langle (\Delta J)^{2}\rangle\over \Delta t}P\biggr
)-{\partial\over\partial J}\biggl (P{\langle
\Delta J\rangle\over \Delta t}\biggr ),
\end{equation}
with
\begin{equation}
\label{ta6}{\langle (\Delta J)^{2}\rangle\over 2 \Delta t}=D, \qquad
{\langle \Delta J\rangle\over \Delta t}={d D\over d
J}+\eta\equiv F^{friction}.
\end{equation}
Substituting Eqs. (\ref{ta3}) and (\ref{ta4}) in Eq. (\ref{ta6}) and using an integration by parts, we find that the diffusion and friction coefficients are given by
\begin{eqnarray}
\label{ta7}{\langle (\Delta J)^{2}\rangle\over 2 \Delta t}=2\pi^2 m\int dJ_1 f_1\sum_{n,n_1}  |A_{n,n_1}(J,J_1)|^2  n^2 \delta(n\Omega(J)-n_1\Omega(J_1)),
\end{eqnarray}
\begin{eqnarray}
{\langle \Delta J\rangle\over \Delta t}=2\pi^2 m\int dJ_1 f_1\sum_{n,n_1} \left ( n\frac{\partial}{\partial J}-n_1\frac{\partial}{\partial J_1}\right ) |A_{n,n_1}(J,J_1)|^2  n \delta(n\Omega(J)-n_1\Omega(J_1)).\nonumber\\
\label{ta8}
\end{eqnarray}
These expressions can be obtained directly from the Hamiltonian
equations of motion by making a systematic expansion of the trajectory
of the particles in powers of $1/N$ in the limit $N\rightarrow
+\infty$ \cite{valageas}.

Let us consider particular cases.

$\bullet$ If we consider a distribution of field particles $f(J_1)$
such that there is no resonance: $n\Omega(J)\neq n_1\Omega(J_1)$ for any couple
$(n,J)\neq (n_1,J_1)$, we first get
\begin{eqnarray}
D=2\pi^2 m\int dJ_1 \sum_{n}  |A_{n,n}(J,J_1)|^2  n^2 \delta(n(\Omega(J)-\Omega(J_1)))
 f_1.
\label{ta9}
\end{eqnarray}
Then, using the identity
$\delta(n(\Omega-\Omega_1))=\delta(J-J_1)/|n\Omega'(J)|$, we find
that
\begin{eqnarray}
\label{ta10}
D(J)=2\pi^2 m f(J)\sum_{n}  |n|  \frac{|A_{n,n}(J,J)|^2}{|\Omega'(J)|}.
\end{eqnarray}
Similarly
\begin{eqnarray}
\eta=2\pi^2 m f'(J)\sum_{n}  |n|  \frac{|A_{n,n}(J,J)|^2}{|\Omega'(J)|}=D(J)\frac{d\ln f}{dJ}.
\label{ta11}
\end{eqnarray}
Using Eq. (\ref{ta6}), we obtain
\begin{equation}
\label{ta12}
{\langle \Delta J\rangle\over \Delta t}=D'(J)+D(J)\frac{d\ln f}{dJ}.
\end{equation}
Finally, the Fokker-Planck equation (\ref{ta2}) can be written
\begin{equation}
\label{ta13}{\partial P\over\partial
t}={\partial\over\partial J}\biggl\lbrack  D(J)\biggl
({\partial P\over\partial J}-P\frac{d\ln f}{dJ}\biggr
)\biggr\rbrack,
\end{equation}
with a diffusion coefficient given by Eq. (\ref{ta9}).

$\bullet$ If the field particles are at statistical equilibrium
(thermal bath), their distribution  is the Boltzmann distribution
\begin{eqnarray}
f({J}_1)=A e^{-\beta m\epsilon(J_1)},\label{ta14}
\end{eqnarray}
where $\epsilon(J)$ is the individual energy of the orbit with action $J$. Using $d\epsilon/dJ=\Omega(J)$, we find that
\begin{eqnarray}
\frac{df_1}{dJ_1}=-\beta m
f_1\frac{d\epsilon_1}{dJ_1}=-\beta m f_1 \Omega(J_1).
\label{ta15}
\end{eqnarray}
Substituting this relation in Eq. (\ref{ta4}), we obtain
\begin{eqnarray}
\eta=-2\pi^2 m^2\beta\int dJ_1 \sum_{n,n_1}  |A_{n,n_1}(J,J_1)|^2  n \delta(n\Omega(J)-n_1\Omega(J_1))
 n_1 f_1 \Omega(J_1).\nonumber\\
\label{ta16}
\end{eqnarray}
Using the $\delta$-function to replace $n_1\Omega(J_1)$ by $n\Omega(J)$ and  comparing the resulting
expression with Eq. (\ref{ta3}), we finally get
\begin{eqnarray}
\eta=-D\beta m \Omega(J), \label{ta17}
\end{eqnarray}
which is the appropriate Einstein relation for our problem.
For a thermal bath, using Eq. (\ref{ta17}), the Fokker-Planck equation
(\ref{ta2}) can be written
\begin{equation}
\label{ta18}{\partial P\over\partial
t}={\partial\over\partial J}\biggl\lbrack  D(J)\biggl
({\partial P\over\partial J}+\beta m P\Omega(J)\biggr
)\biggr\rbrack,
\end{equation}
where $D(J)$ is given by Eq. (\ref{ta3}) with Eq. (\ref{ta14}).

\section{Discussion and conclusion}
\label{conclusion}

The kinetic theory of systems with long-range interactions
leads to the following scenario:

1. The first stage of the dynamical evolution is described by the
Vlasov equation. Due to violent relaxation and phase mixing, the
coarse-grained distribution function $\overline{f}({\bf r},{\bf v},t)$
converges towards a steady state $\overline{f}({\bf r},{\bf v})$ which
is a dynamically stable (robust) steady state of the Vlasov
equation. The convergence takes a few dynamical times $t_D$
independent of $N$. The fine-grained distribution $f({\bf r},{\bf
v},t)$ develops filaments at smaller and smaller scales and does not
achieve a steady state (presumably). This collisionless relaxation
explains the rapid emergence of coherent structures in stellar
dynamics (galaxies), two-dimensional turbulence (jets and vortices)
and in the HMF model (quasistationary states).

2. If the system mixes efficiently, the QSS is given by the
Lynden-Bell statistical theory of the Vlasov equation. This is the
usual Boltzmann approach taking into account the specificities of the
Vlasov equation (Casimir constraints). Therefore, the Lynden-Bell
prediction depends on the {\it details} of the initial condition, not
only on the mass and energy. However, violent relaxation can be incomplete:
the system may not mix well and the collisionless relaxation may not
be ergodic.  In that case, other distributions can emerge. Among them,
the Tsallis distributions (polytropes) seem to play a particular role
in certain situations.  However, they are not universal
attractors. Other distributions can emerge as well but they are
difficult to predict since they depend on the dynamics and on the
efficiency of mixing.

3. On longer timescales, the system evolves under the effect of ``collisions'',
i.e. correlations due to graininess (finite $N$) effects. This gives rise to
a collision term in the r.h.s. of the Vlasov equation. In principle, the
collisional relaxation is towards the Boltzmann statistical equilibrium state
(for stellar systems, the relaxation towards Boltzmann equilibrium is hampered
by the problems of evaporation and gravothermal catastrophe). The collisional
relaxation time $t_{R}(N)$ depends on the number of particles $N$ and
diverges rapidly as $N\rightarrow +\infty$.

4. Due to the huge timescale separation between the dynamical time
$t_D$ and the relaxation time $t_{R}(N)$, the distribution function
$f({\bf r},{\bf v},t)$ passes by a sequence of quasi equilibrium
states which are quasi stationary states of the Vlasov equation slowly
evolving in time due to finite $N$ effects (graininess). In stellar
dynamics, this collisional evolution is described by the
orbit-averaged-Fokker-Planck equation.

This scenario was put forward by H\'enon
\cite{henon,henonVR} in the context of stellar dynamics and 
further discussed and improved by many authors. It has become the ``standard
scenario'' of stellar dynamics \cite{spitzerbook,bt}.  A similar
scenario was proposed by Chavanis \cite{kin,houches} for
two-dimensional point vortices, by developing the analogy between the
dynamics of stellar systems and point vortices. Inspired by these
results, this scenario was also considered by Yamaguchi {\it
et al.} \cite{yamaguchi} in relation to the HMF model. These authors understood
many important aspects of the dynamics. However, we would like to
complete some of their arguments in the light of more recent
observations: (i) Yamaguchi {\it et al.} \cite{yamaguchi} argue that,
during the collisional evolution, the system remains always close to a
{\it stable} steady state of the Vlasov equation.  Recently, the
numerical study of Campa {\it et al.}
\cite{campat} has demonstrated that, during the collisional evolution,
the distribution function $f(v,t)$ can be well-represented by a
Tsallis \cite{tsallis} distribution (polytrope) with a time dependent
index $q(t)$ until it becomes Vlasov {\it unstable} and triggers a
dynamical phase transition from the non magnetized (homogeneous) phase
to the magnetized (inhomogeneous) phase. Therefore, in that case,
the relaxation involves a {\it dynamical instability} \cite{campat}. In
fact, at the verge of the dynamical phase transition, the imaginary
part of the complex pulsation $\omega_i$ becomes zero and it is
possible that the system remains {\it marginally stable} in the
subsequent evolution (this would be an interesting point to
check). (ii) Yamaguchi {\it et al.}
\cite{yamaguchi} reject the possibility of Tsallis $q$-distributions
\cite{tsallis} although they seem to play an important role in the HMF
model as discussed in
\cite{cc}. (iii) They argue that the relaxation time scales
algebraically like $N^{1.7}$ although, as we shall see below, the
picture is more complex.

We also emphasize that the above scenario is
not the only possibility and that, as reviewed in \cite{incomplete,paper3}, the
evolution can be more complex with the formation of phase space holes, periodic
solutions, oscillations, vortex crystals etc.

We would like now to use the kinetic theory presented in the present paper
to interpret the results of numerical simulations that have been obtained
in different contexts. We briefly review these numerical results, then
provide new interpretations that can open interesting directions of research.

{\it Stellar systems}: the scenario proposed by H\'enon
\cite{henon,henonVR} has been studied  by several authors. The theory
of violent relaxation developed by Lynden-Bell \cite{lb} explains the
isothermal core of elliptical galaxies without recourse to collisions
that operate on a much longer timescale.  However, his theory does not
explain the structure of the halo that results from {\it incomplete
relaxation}. The velocity distribution in the halo is radially
anisotropic and the density profile decreases like $r^{-4}$ (instead
of the $r^{-2}$ isothermal law). Theories of incomplete relaxation
have been elaborated by Stiavelli \& Bertin \cite{sb}, Hjorth \&
Madsen \cite{hm} and Chavanis {\it et al.} \cite{csr}.  Numerical
simulations of violent relaxation have been made by van Albada
\cite{albada} who compared his results with the phenomenological de
Vaucouleur $R^{1/4}$ law. In the collisional regime, the evolution is
first governed by the evaporation of high energy stars. Numerical
simulations \cite{spitzerbook} show that the system reaches a quasi
stationary state close to the Michie-King distribution that slowly
evolves in amplitude due to the evaporation as the system loses mass
and energy. The density follows the isothermal law $r^{-2}$ in the
central region (with a core of almost uniform density) and decreases
like $r^{-7/2}$ in the halo (that has anisotropic velocity
distribution).  At some point of the evolution, when the density
contrast between the core and the halo becomes sufficiently high, the
system undergoes a thermodynamical instability related to the Antonov
\cite{antonov} instability, and the gravothermal catastrophe
\cite{lbw} takes place. Core collapse has been studied numerically by
Cohn \cite{cohn} using the orbit-averaged-Fokker-Planck equation. He
finds that the collapse is self-similar, that the central density
becomes infinite in a finite time, and that the density behaves like
$r^{-2.23}$.  When the system is confined in a box, instead of being
free to expand, Taruya \& Sakagami \cite{ts} observe numerically that
the evolution follows a sequence of $q$-distributions (polytropes)
with a time dependent index $q(t)$ until an instability develops (that
we interprete as a Vlasov dynamical instability in \cite{campat}). We
think that a good understanding of stellar dynamics has been reached
by astrophysicists for a long time.

{\it Point vortices}: the scenario proposed by Chavanis
\cite{kin,houches}, by analogy with stellar dynamics, has been tested
numerically by Kawahara \& Nakanishi \cite{kawahara,kawahara2} in two
very interesting papers.  We would like to complete their discussion
in the light of the kinetic theory presented in this paper.  The
kinetic theory developed in \cite{kin,houches,kinvortex} is valid at
the order $O(1/N)$ and, consequently, the natural timescale is $N
t_D$. This is precisely the relaxation time that Kawahara \& Nakanishi
observe \cite{kawahara2}. This seems consistent at first sight with
the kinetic theory. However, if we consider things carefully, the
situation appears to be more complicated. Indeed, we have seen in
Sec. \ref{sec_vw} that the collision term can vanish leading to a
relaxation time larger than $Nt_D$. However, we have indicated that
this cancelation occurs only for axisymmetric systems with monotonic
profile of angular velocity. Now, in the simulations of Kawahara \&
Nakanishi \cite{kawahara2} the system is non-axisymmetric. Indeed, in
most cases, the phase of violent relaxation leads to a vortex crystal
(see their Figs 2.b and 7). Therefore, we cannot use the axisymmetric
kinetic equation (\ref{vw19}) and we must come back to the more
general kinetic equation (\ref{vw2}). Little is known concerning this
complicated equation but we expect that it increases the entropy and
conserves energy and circulation \cite{kinvortex}. It is not clear
whether this equation rigorously converges towards the Boltzmann
distribution but, since the entropy increases, we deduce that the
system has the {\it tendency} to approach the Boltzmann distribution
(it becomes more mixed). This is because there are additional
resonances with respect to a purely axisymmetric evolution. Since the
proper timescale of this kinetic equation is $N t_D$, the kinetic
theory of {\it non-axisymmetric flows} may explain the numerical
observations of Kawahara \& Nakanishi \cite{kawahara2}. There remains,
however, a difficulty. Indeed, in their Fig. 2a, violent relaxation
leads to an apparently axisymmetric distribution with a presumably
monotonically decreasing profile of angular velocity. In that case,
there should be no resonance and the relaxation time should be larger
than $N t_D$ contrary to what is observed numerically. We speculate
that, in this case, the system is not really axisymmetric but that
there exists ``structures'' in the flow (analogous to Dawson's
vortices in neutral plasmas under a strong magnetic field
\cite{dawson}).  This claim is corroborated by the fact that Kawahara
\& Nakanishi \cite{kawahara2} observe anomalous diffusion which means
that the evolution of the flow is more complex than it seems. These
results have been confirmed recently by Yoshida \cite{yoshida} who
showed that the mean square displacements exhibit a universal time
dependence $\langle \lbrack r(t)-r(0)\rbrack^2\rangle\sim t^\mu$ with
$\mu=1.75\pm 0.1$. These authors attribute anomalous diffusion to
occasional long jumps of the particles convected by long living large
vortices. If there exists such ``structures'', they may induce new
resonances and explain the relaxation towards Boltzmann equilibrium on
a timescale $Nt_D$.  This may also imply that there is no
contradiction with the kinetic theory concerning the diffusion
process. The diffusion coefficient (\ref{tv9}) and the Fokker-Planck
equation (\ref{tv12}) derived in \cite{kin,houches} are only valid
for axisymmetric flows without ``structures''.  In more general
cases, the diffusion equation is given by Eq. (142) of
\cite{kinvortex} which involves delocalization in space and time. It
is plausible that these non ideal effects lead to anomalous diffusion
for short times.  Normal diffusion should be recovered for larger
times (anomalous diffusion could also be a finite size effect and it
could disappear for large $N$).  As indicated by Kawahara \& Nakanishi
\cite{kawahara2}, this is precisely what happens for neutral plasmas
under a strong magnetic field in relation to the Taylor-McNamara
\cite{tm} diffusion.  Of course, this scenario demands to be confirmed
but it seems plausible enough to reconcile theory and
observations. More work remains to be done in order to better
understand these results. In this respect, we may note the recent
study of Sano \cite{sano} on the kinetic theory of point vortices. He
confirmed the results of Chavanis \cite{kin,kinvortex} (leading to
Landau-type kinetic equations) and derived a kinetic equation
corresponding to the Lenard-Balescu equation for plasmas using the
theory of the Fredholm integral equation. We may also note that the
exponential tail of the velocity distribution observed by Kawahara \&
Nakanishi
\cite{kawahara2} is consistent with the results reported by Chavanis
\& Sire \cite{cs2} when the vortices have a finite core, which is
equivalent to putting a restriction on the point vortex distance in
the initial state. Therefore, this observation (exponential tail) is
not inconsistent with kinetic theory.

{\it HMF model}: the scenario proposed by Yamaguchi {\it et al.}
\cite{yamaguchi} has been checked numerically by these authors and by
Campa {\it et al.} \cite{campa,campat}.  The relaxation time has been
discussed by Bouchet \& Dauxois \cite{bd} and Chavanis \cite{paper3}
in the light of the kinetic theory. These authors argue that, for
spatially homogeneous systems, the relaxation time is larger than $N
t_D$ since there is no resonance.  Bouchet \& Dauxois \cite{bd} claim
that this result is consistent with the timescale $N^{1.7}$
numerically observed by Yamaguchi {\it et al.}
\cite{yamaguchi}. Chavanis \cite{paper3} claims that this is
consistent with the timescale $e^N$ observed by Campa {\it et al.}
\cite{campa}.  We would like to reconsider and complete their
discussions. In the simulation of Yamaguchi {\it et al.}
\cite{yamaguchi}, that is performed for subcritical energies $U<U_c$,
the system does {\it not} remain spatially homogeneous so that the
homogeneous kinetic theory cannot be advocated until the
end. Alternatively, Campa {\it et al.} \cite{campa} perform
supercritical simulations $U>U_c$, in which the system remains
permanently spatially homogeneous, and find that the relaxation time
scales like $e^N$. Therefore, when the system is homogeneous the
relaxation time is exponentially long! This is in agreement with the
kinetic theory but this shows that the relaxation is not due to just
three-body, four-body, $n$-body...  correlations (as we could naively
believe), that would imply a relaxation time scaling as $N^2$, $N^3$,
$N^{n-1}$... All the correlations functions seem to matter!
Therefore, the kinetic theory of spatially homogeneous systems cannot
apparently be truncated at some power of $1/N$. Using the analogy
between the point vortex model and the HMF model, we may conjecture
that the relaxation time for an axisymmetric distribution of point
vortices with monotonic profile of angular velocity scales like $e^N$
(although the possibility that such distributions never achieve
statistical equilibrium cannot be rejected). When the system is
spatially inhomogeneous, the relaxation time is reduced with respect
to $e^N$. This leads to a reinterpretation of the numerical
results. Yamaguchi {\it et al.} \cite{yamaguchi} performed two types
of simulations: (i) the first one, corresponding to unstable
stationary Vlasov states, leads to a spatially inhomogeneous
evolution. In that case, the collisional relaxation times scales like
$N$ which is the proper scaling of the kinetic theory for spatially
inhomogeneous systems. Indeed, there exists many resonances that
``push'' the system towards the Boltzmann distribution (see
Sec. \ref{sec_angleaction}). The question whether the system really
reaches the Boltzmann distribution, or a distribution {\it close} to
it, remains open. (ii) The second, which starts from Vlasov stable
stationary states, leads to spatially homogeneous structures. In that
case, there is no resonance and we should expect a relaxation time
scaling as $e^N$. However, Campa {\it et al.}
\cite{campat} argue that the relaxation towards Boltzmann equilibrium
involves a dynamical phase transition that is triggered when the
homogeneous phase becomes Vlasov unstable. In that case, the system
becomes spatially inhomogeneous and resonances appear. This
considerably accelerate the relaxation with respect to the exponential
scaling and leads to a non trivial (but probably non universal)
scaling $N^{1.7}$. Therefore, in that case, the timescale cannot be
solely understood in terms of the kinetic theory of homogeneous
systems, although this theory explains why the scaling is larger than
$N$.  Morita \& Kaneko \cite{mk}, considering an initial condition
with $U<U_c$ and $M_0=1$, found a relaxation time $t_R\sim Nt_D$. In
their simulations, the system is always spatially inhomogeneous (the
magnetization has an oscillatory behavior) and the relaxation time is
consistent with the natural scaling of the kinetic theory for
inhomogeneous distributions. Their results may be consistent with the
general kinetic equations (\ref{sw2}) and (\ref{ses14}) although this is
of course difficult to check.  Let us finally consider the early
simulation of Latora {\it et al.}
\cite{latora} which starts from an out-of-equilibrium initial state
with magnetization $M_0=1$. Violent relaxation leads the system to a
spatially homogeneous QSS which is followed by a collisional
relaxation towards the Boltzmann distribution. They find a relaxation
time scaling like $Nt_D$. This timescale seems to be inconsistent with
the kinetic theory of spatially homogeneous systems. However, as in
the case of 2D point vortices, we speculate that there exists
``phase-space structures'' that make the system spatially
inhomogeneous and induce additional resonances. These inhomogeneities
may reduce the timescale of relaxation to the natural timescale
$Nt_D$. These ``phase-space structures'' are indeed present in the
simulations of Latora {\it et al.} \cite{latora} and they seem to be
responsible for anomalous diffusion in a way similar to what happens
in 2D point vortex dynamics \cite{kawahara2,yoshida}. Again, this scenario
demands to be confirmed but it seems plausible enough to reconcile
theory and observations. We must, however, be very careful because
these striking features (phase-space structures and anomalous
diffusion) may be due to finite size effects \cite{ma,prelb} and
disappear for $N\rightarrow +\infty$.  Also, diffusion may be
anomalous for short times and normal for large times. This may be a
way to reconcile the approaches of Rapisarda \& Pluchino \cite{rp} and
Bouchet \& Dauxois \cite{bd} who studied temporal correlation
functions and anomalous diffusion. Indeed, their studies are based on
very different arguments: Bouchet \& Dauxois \cite{bd} assume that the
background distribution is spatially homogeneous and use standard
kinetic theory based on Fokker-Planck equations.  Alternatively,
Rapisarda \& Pluchino \cite{rp} assume that anomalous diffusion is due
to the presence of phase-space structures and use nonlinear
Fokker-Planck equations.  We believe that these two approaches are not
antagonistic but that they apply to different regimes or, maybe,
different numbers of particles.

In conclusion, we have developed a kinetic theory of systems with
long-range interactions based on the generalized (and not well-known)
kinetic equations (\ref{sw2}), (\ref{vw2}) and (\ref{ses1}). A
specificity of our approach is to remain in physical space and take
into account spatial inhomogeneity which is the norm for systems with
long-range interactions. This is the main difference with kinetic
theories of systems with short-range interactions and neutral plasmas
that are spatially homogeneous. Interestingly, the homogeneous kinetic
equation can be easily derived from the generalized kinetic equation
with simple calculations. This provides a new derivation of
Landau-type equations and also offers a unified description of the
kinetic theory of stellar systems, 2D point vortices and the HMF
model. As an application, we have argued that the relaxation time of
spatially inhomogeneous 1D systems (or non axisymmetric 2D flows)
should scale as $N t_D$ which is the natural timescale of the kinetic
theory. Indeed, spatially inhomogeneous systems involve many
resonances that allow the relaxation (or the approach) of the system
towards Boltzmann equilibrium. By contrast, for one dimensional
homogeneous systems (or axisymmetric flows with monotonic profile of
angular velocity), the resonances disappear and the relaxation
timescale is considerably increased.  We have speculated that, when
the system remains permanently homogeneous, the natural timescale is
$e^N t_D$. In the HMF model, this corresponds to supercritical
energies considered by Campa {\it et al.}
\cite{campa}. If the QSS resulting from violent relaxation is
homogeneous but, later evolving under the development of correlations,
undergoes at some point a dynamical phase transition (Vlasov
instability) from the homogeneous phase to the inhomogeneous phase, the
relaxation time should be intermediate between $N t_D$ and $e^N
t_D$. For example, it can scale algebraically as $N^\delta t_D$. This
corresponds to subcritical energies considered by Yamaguchi {\it et
al.} \cite{yamaguchi}. They find $\delta=1.7$ but this scaling may not
be universal. We have motivated our arguments by a detailed comparison
with existing numerical simulations and found interesting
agreements. This shows that kinetic theory can be a good approach to
understand the dynamics and thermodynamics of systems with long-range
interactions but that many works remain to be done to obtain a
complete description of spatially inhomogeneous systems.

\appendix

\section{Calculation of $K^{\mu\nu}$}
\label{sec_Kmunu}

In $d=3$, we introduce a spherical system of coordinates where the $z$ axis is taken in the direction of ${\bf w}$. Then
\begin{eqnarray}
K^{\mu\nu}=\pi (2\pi)^3 m^2 \int_{0}^{+\infty} k^2dk\int_{0}^{\pi}\sin\theta\, d\theta \int_{0}^{2\pi}d\phi\,   k^{\mu}k^{\nu} \delta(kw\cos\theta) \hat{u}(k)^2.\label{k1}
\end{eqnarray}
Using $k_{x}=k\sin\theta\cos\phi$, $k_y=k\sin\theta\sin\phi$ and
$k_z=k\cos\theta$, it is easy to see that only $K^{xx}$, $K^{yy}$ and
$K^{zz}$ can be non-zero. The other components of the matrix
$K^{\mu\nu}$ vanish by symmetry. Furthermore
\begin{eqnarray}
K^{xx}=K^{yy}=\pi^2 (2\pi)^3 m^2 \int_{0}^{+\infty} k^2\, dk\int_{0}^{\pi}\sin\theta\, d\theta\,   k^{2} \delta(kw\cos\theta) \hat{u}(k)^2 \sin^{2}\theta.\nonumber\\
\label{k2}
\end{eqnarray}
Using the identity $\delta(\lambda x)=\frac{1}{|\lambda|}\delta(x)$, we get
\begin{eqnarray}
K^{xx}=K^{yy}=\pi^2 (2\pi)^3 m^2 \frac{1}{w}\int_{0}^{+\infty} k^3  \hat{u}(k)^2 \, dk\int_{0}^{\pi}\sin^3\theta \delta(\cos\theta) \, d\theta.
\label{k4}
\end{eqnarray}
With the change of variables $t=\cos\theta$, we obtain
\begin{eqnarray}
K^{xx}=K^{yy}=\pi^2 (2\pi)^3 m^2 \frac{1}{w}\int_{0}^{+\infty} k^3  \hat{u}(k)^2 \, dk \int_{-1}^{+1} (1-t^2) \delta(t) \, dt,
\label{k5}
\end{eqnarray}
so that finally
\begin{eqnarray}
K^{xx}=K^{yy}=8 \pi^5  m^2 \frac{1}{w}\int_{0}^{+\infty} k^3  \hat{u}(k)^2 \, dk.
\label{k6}
\end{eqnarray}
On the other hand
\begin{eqnarray}
K^{zz}=2\pi^2 (2\pi)^3 m^2 \frac{1}{w} \int_{0}^{+\infty} k^3 \hat{u}(k)^2\, dk \int_{0}^{\pi}\sin\theta\cos^{2}\theta \delta(\cos\theta)  \, d\theta=0.
\label{k7}
\end{eqnarray}
In conclusion, we obtain
\begin{eqnarray}
K^{\mu\nu}=\frac{K}{w}\left (\delta^{\mu\nu}-\frac{w^{\mu}w^{\nu}}{w^2}\right ),
\label{k8}
\end{eqnarray}
with
\begin{eqnarray}
K=8 \pi^5  m^2 \int_{0}^{+\infty} k^3  \hat{u}(k)^2 \, dk.
\label{k9}
\end{eqnarray}

In $d=2$, we introduce a polar system of coordinates where the $x$ axis is taken in the direction of ${\bf w}$. Then
\begin{eqnarray}
K^{\mu\nu}=\pi (2\pi)^2 m^2 \int_{0}^{+\infty} k \, dk\int_{0}^{2\pi} d\theta \,    k^{\mu}k^{\nu} \delta(kw\cos\theta) \hat{u}(k)^2.\label{k10}
\end{eqnarray}
Using $k_{x}=k\cos\theta$, $k_y=k\sin\theta$, it is easy to see that only $K^{xx}$ and $K^{yy}$ can be non-zero. The other components of the matrix
$K^{\mu\nu}$ vanish by symmetry. Furthermore
\begin{eqnarray}
K^{yy}=\pi (2\pi)^2 m^2 \int_{0}^{+\infty} k\, dk\int_{0}^{2\pi} d\theta\, k^2 \delta(kw\cos\theta) \hat{u}(k)^2  \sin^{2}\theta,
\label{k11}
\end{eqnarray}
or, equivalently,
\begin{eqnarray}
K^{yy}=2\pi (2\pi)^2 m^2 \frac{1}{w}\int_{0}^{+\infty} k^2\, dk\, \hat{u}(k)^2 \int_{0}^{\pi} d\theta\, \delta(\cos\theta)  \sin^{2}\theta.
\label{k12}
\end{eqnarray}
With the change of variables $t=\cos\theta$, we obtain
\begin{eqnarray}
K^{yy}=2\pi (2\pi)^2 m^2 \frac{1}{w}\int_{0}^{+\infty} k^2 \hat{u}(k)^2 \, dk \int_{-1}^{+1} \sqrt{1-t^2} \delta(t) \, dt,
\label{k13}
\end{eqnarray}
so that finally
\begin{eqnarray}
K^{yy}=8 \pi^3  m^2 \frac{1}{w}\int_{0}^{+\infty} k^2  \hat{u}(k)^2 \, dk.
\label{k14}
\end{eqnarray}
On the other hand
\begin{eqnarray}
K^{xx}=\pi (2\pi)^2 m^2 \frac{1}{w}\int_{0}^{+\infty} k^2\, dk\, \hat{u}(k)^2 \int_{0}^{2\pi} d\theta\, \delta(\cos\theta)  \cos^{2}\theta=0.
\label{k15}
\end{eqnarray}
In conclusion, we obtain
\begin{eqnarray}
K^{\mu\nu}=\frac{K}{w}\left (\delta^{\mu\nu}-\frac{w^{\mu}w^{\nu}}{w^2}\right ),
\label{k16}
\end{eqnarray}
with
\begin{eqnarray}
K=8 \pi^3  m^2 \int_{0}^{+\infty} k^2  \hat{u}(k)^2 \, dk.
\label{k17}
\end{eqnarray}

\section{Calculation of $\chi(r,r_1)$}
\label{sec_chi}

For the ordinary potential $u(|{\bf r}-{\bf
r}_{1}|)=-\frac{1}{2\pi}\ln |{\bf r}-{\bf r}_{1}|$, we have
\begin{eqnarray}
u(r,r_1,\phi)=-\frac{1}{4\pi}\ln (r^2+r_1^2-2rr_1\cos\phi),
\label{c1}
\end{eqnarray}
where $\phi=\theta-\theta_1$.
Its Fourier transform on the angles can be written
\begin{eqnarray}
\hat{u}_{m}(r,r_1)=-\frac{1}{4\pi^2}\int_{0}^{\pi}\ln(1-\lambda\cos\phi)\cos(m\phi)\, d\phi-\frac{\delta_{m0}}{4\pi}\ln(r^2+r_1^2),
\label{c2}
\end{eqnarray}
with $\lambda=\frac{2rr_1}{r^2+r_1^2}$. For $m\neq 0$, we have the identity
\begin{eqnarray}
\int_{0}^{\pi}\ln\left (1-\lambda\cos\phi\right )\cos(m\phi)\, d\phi=-\frac{\pi}{|m|}\left (\frac{1}{\lambda}-\sqrt{\frac{1}{\lambda^2}-1}\right )^{|m|}.
\label{c3}
\end{eqnarray}
Therefore
\begin{eqnarray}
\hat{u}_{m}(r,r_1)=\frac{1}{4\pi |m|}\left (\frac{r^2+r_1^2-|r^2-r_1^2|}{2 r r_1}\right )^{|m|}=\frac{1}{4\pi |m|}\left (\frac{r_{<}}{r_>}\right )^{|m|},
\label{c4}
\end{eqnarray}
where $r_<$ (resp. $r_>$) is the min (resp. max) or $r$ and $r_1$. For $m=0$, we have the identity
\begin{eqnarray}
\int_{0}^{\pi}\ln\left (1-\lambda\cos\phi\right )\, d\phi=\pi\ln(1+\sqrt{1-\lambda^2})-\pi\ln 2.
\label{c5}
\end{eqnarray}
Therefore
\begin{eqnarray}
\hat{u}_{0}(r,r_1)=-\frac{1}{4\pi}\ln\left (\frac{r^2+r_1^2+|r^2-r_1^2|}{2}\right )=-\frac{1}{2\pi}\ln r_>.
\label{c6}
\end{eqnarray}
Combining the previous results, the potential of interaction can be written
\begin{eqnarray}
u(r,r_1,\phi)=-\frac{1}{2\pi}\ln r_>+\frac{1}{4\pi}\sum_{m\neq 0}\frac{1}{|m|}\left (\frac{r_<}{r_>}\right )^{|m|} e^{im\phi}.
\label{c7}
\end{eqnarray}
The function defined by Eq. (\ref{vw20}) takes the form
\begin{eqnarray}
\chi(r,r_1)=\frac{1}{8\pi^2}\sum_{m=1}^{+\infty}\frac{1}{m}\left (\frac{r_<}{r_>}\right )^{2m}=-\frac{1}{8\pi^2}\ln\left\lbrack 1-\left (\frac{r_<}{r_>}\right)^2\right\rbrack.
\label{c8}
\end{eqnarray}
In particular
\begin{eqnarray}
\chi(r,r)=\frac{1}{8\pi^2}\sum_{m=1}^{+\infty}\frac{1}{m}=\frac{1}{8\pi^2}\ln\Lambda,
\end{eqnarray}
where $\ln\Lambda$ is a  Coulombian logarithm that has
to be regularized appropriately (see, e.g. \cite{cl}).

\vskip1cm

\noindent{\bf References}




\end{document}